\documentclass[letterpaper,twocolumn,10pt]{usetex-v1}

\usepackage[english]{babel}
\usepackage{times}
\usepackage{graphicx}
\usepackage{algorithm}
\usepackage{algpseudocode}
\usepackage{color}
\usepackage{endnotes}
\usepackage{listings}
\usepackage{makecell}
\usepackage{multirow}
\usepackage{url}
\usepackage{cite}
\usepackage[colorlinks=false,linkcolor=blue,citecolor=black,pdfborder={0 0 0}]{hyperref}
\usepackage{breakurl}
\usepackage{enumitem}
\usepackage{xcolor}
\usepackage{xspace}
\usepackage[kerning,spacing]{microtype}
\usepackage[font=small,labelfont=bf]{caption}
\usepackage{amsmath}
\usepackage[export]{adjustbox}
\usepackage{todonotes}

\usepackage{mathptmx}
\DeclareMathAlphabet{\mathcal}{OMS}{cmsy}{m}{n}
\DeclareMathAlphabet{\mathrm}{OT1}{bch}{m}{n}
\DeclareMathAlphabet{\mathit}{OT1}{bch}{m}{it}

\setitemize{noitemsep,topsep=0pt,parsep=0em,partopsep=0pt}

\newcommand{\sysname}{\textsf{SepBIT}\xspace}
\renewcommand{\paragraph}[1]{{\smallskip\noindent\bf #1}}

\begin{document}

\title{Separating Data via Block Invalidation Time Inference \\
for Write Amplification Reduction in Log-Structured Storage\thanks{The
conference version of this paper appeared at FAST'22 \cite{wang22}.}}
\author{Qiuping Wang$^{1,2}$, Jinhong Li$^1$, Patrick P. C. Lee$^1$, Tao Ouyang$^2$, Chao Shi$^2$, Lilong Huang$^2$\\
  $^1${\em The Chinese University of Hong Kong} \ \
  $^2${\em Alibaba Group}
\vspace{-9pt}
}

\maketitle

\begin{abstract}
Log-structured storage has been widely deployed in various domains of storage
systems, yet its garbage collection incurs write amplification
(WA) due to the rewrites of live data.  We show that there exists an
optimal data placement scheme that minimizes WA using the future knowledge
of block invalidation time (BIT) of each written block, yet it is infeasible
to realize in practice.  We propose a novel data placement algorithm for
reducing WA, \sysname, that aims to infer the BITs of written blocks from
storage workloads and separately place the blocks into groups with
similar estimated BITs.  We show via both mathematical and production trace
analyses that \sysname effectively infers the BITs by leveraging the write
skewness property in practical storage workloads.  Trace analysis and
prototype experiments show that \sysname reduces WA and improves I/O
throughput, respectively, compared with state-of-the-art data placement
schemes. \sysname is currently deployed to support the log-structured block
storage management at Alibaba Cloud.
\end{abstract}

\section{Introduction}
\label{sec:intro}

Modern storage systems adopt the {\em log-structured} design
\cite{rosenblum92} for high performance.  Examples include flash-based
solid-state drives (SSDs) \cite{agrawal08,chen09}, file systems
\cite{rosenblum92,hartman93,seltzer93,min12,lee15}, key-value
stores \cite{oneil96, lu16},
table stores \cite{chang06}, storage management
\cite{balakrishnan13}, in-memory storage \cite{rumble14}, RAID arrays
\cite{kim19swan}, and cloud block services \cite{xu19}.  Log-structured storage
transforms random write requests into sequential disk writes in an append-only
log, so as to reduce disk seek overhead and improve write performance.  It also
brings various advantages in addition to high write performance, such as
improved flash endurance in SSDs \cite{lee15}, unified abstraction for building
distributed applications \cite{chang06,balakrishnan13}, efficient memory
management in in-memory storage \cite{rumble14}, and load balancing in cloud
block storage \cite{xu19}.  Recent advances in zoned storage
\cite{zonedstorage,bjorling21} also advocate the adoption of log-structured
storage based on append-only interfaces for scalable performance. 

The log-structured design writes live data blocks to the append-only log
without modifying existing data blocks in-place, so it regularly performs {\em
garbage collection (GC)} to reclaim the free space of stale blocks. GC works
by reading a segment of blocks, removing any stale blocks, and writing back
the remaining live blocks. The repeated writes of live blocks lead to {\em
write amplification (WA)}.
  They not only incur I/O interference to foreground workloads
\cite{kim19swan}, but also lead to reduced flash lifespans and unnecessary
power consumption in data centers.

Mitigating WA in log-structured
storage has been a well-studied topic in the literature 
(\S\ref{sec:related}).  In particular, a large body of studies
focuses on designing {\em data placement} strategies by properly placing blocks
in separate groups.  He {\em et al.} \cite{he17} point out that a data
placement scheme should group blocks by the {\em block invalidation time (BIT)}
(i.e., the time when a block is invalidated by a live block; a.k.a. the death
time \cite{he17}) to achieve the minimum WA.  However, without
obtaining the future knowledge of the BIT pattern, how to design an optimal
data placement scheme with the minimum WA remains an unexplored issue.
Existing temperature-based data placement schemes that group blocks by block
temperatures (e.g., write/update frequencies)
\cite{chiang99,min12,stoica13,shafaei16,yang17,kremer19,yang19} are arguably
inaccurate to capture the BIT pattern and fail to effectively group the blocks
with similar BITs \cite{he17}.

We propose \sysname, a novel data placement scheme that aims for
the minimum WA in log-structured storage.  It infers the BITs of written
blocks from the underlying storage workloads and separately places the
written blocks into different groups, each of which stores the blocks with
similar {\em estimated} BITs.  Specifically, it builds on the {\em skewed}
write patterns observed in the real-world cloud block storage workloads (e.g.,
Alibaba Cloud \cite{li20} and Tencent Cloud \cite{zhang20osca}). 
It separates the written blocks into {\em user-written blocks} and {\em
GC-rewritten blocks} (defined in \S\ref{subsec:gc}). It further separates
each set of user-written blocks and GC-rewritten blocks by inferring the BIT
of each block, so as to perform fine-grained separation of blocks into groups
with similar estimated BITs.  We summarize our contributions below:
\begin{itemize}[leftmargin=*]
\item
We first design an ideal data placement strategy that has the minimum WA in
log-structured storage, based on the (impractical) assumption of having the
future knowledge of BITs of written blocks. Our analysis not only motivates 
how to design a practical data placement scheme that aims to group the written
blocks with similar BITs, but also provides an oracular baseline
for our comparisons.
\item
We design \sysname, which performs fine-grained separation of written blocks
by inferring their BITs from the underlying storage workloads.  We show via
both mathematical and trace analyses that our BIT inference is effective in
skewed workloads.  \sysname also achieves low memory overhead in its indexing
structure for tracking block statistics. 
\item
We evaluate \sysname using real-world cloud block storage workloads at
Alibaba Cloud \cite{li20} and Tencent Cloud \cite{zhang20osca}.  Trace
analysis on both workloads shows that \sysname has the lowest WA compared with
eight state-of-the-art data placement schemes. For example, for the Alibaba
Cloud traces, \sysname reduces the overall WA by 8.6-15.9\% and 9.1-20.2\%
when the Greedy \cite{rosenblum92} and Cost-Benefit
\cite{rosenblum92,rumble14} algorithms are used for segment selection in GC,
respectively.  It also reduces the per-volume WA by up to 44.1\%, compared
with merely separating user-written and GC-rewritten blocks in data placement. 
\item
We prototype a log-structured storage system that supports different data
placement schemes and runs on an emulated zoned storage backend based on ZenFS
\cite{zenfs}.  Our prototype experiments show that \sysname improves I/O
throughput over most volumes due to its efficient WA reduction;
for example, its median throughput is 20\% higher than the second best data
placement scheme.
\end{itemize}

\medskip
\sysname is currently deployed at Alibaba Cloud Enhanced SSDs (ESSDs)
\cite{essd}, which provide cloud block storage services for end-users or
applications.  Each ESSD is a block-level volume (or virtual disk) backed by
flash-based SSD storage, and aims to support low-latency (e.g., around
100$\mu$s) and high-throughput (e.g., up to 1\,M IOPS) I/O access.  ESSDs are
deployed atop Pangu \cite{pang21}, a general distributed storage platform that
provides an append-only write interface.  To be compatible with the
append-only write interface of Pangu, ESSDs adopt the log-structured design
and are abstracted as log-structured storage in our paper.

Our trace analysis scripts and prototype are open-sourced at
\url{http://adslab.cse.cuhk.edu.hk/software/sepbit}.

\section{Background and Motivation}
\label{sec:background}

\subsection{GC in Log-Structured Storage}
\label{subsec:gc}

We consider a log-structured storage system that comprises multiple 
{\em volumes}, each of which is assigned to a user.  Each volume is configured
with a capacity of tens to hundreds of GiB and manages data in an append-only
manner. It is further divided into {\em segments} that are configured with a
maximum size (e.g., tens to hundreds of MiB).  Each segment contains fixed-size
{\em blocks}, each of which is identified by a {\em logical block address
(LBA)} and has a size (e.g., several KiB) that aligns with the underlying disk
drives.  Each block, either from a new write or from an update to an existing
block, is appended to a segment (called an {\em open} segment) that has not yet
reached its maximum size. If a segment reaches its maximum size, the segment
(called a {\em sealed} segment) becomes immutable.  Updating an existing block
is done in an {\em out-of-place} manner, in which the latest version of the
block is appended to an open segment and becomes a {\em valid} block, while the
old version of the block is invalidated and becomes an {\em invalid} block.

Log-structured storage needs to regularly reclaim the space occupied by the
invalid blocks via GC.  A variety of GC policies can be realized, yet we can
abstract a GC policy as a three-phase procedure: 
\begin{itemize}[leftmargin=*]
\item
{\em Triggering}, which decides when a GC operation should be activated. In
this work, we assume that a GC operation is triggered for a volume when its 
{\em garbage proportion (GP)} (i.e., the fraction of invalid blocks among all
valid and invalid blocks) exceeds a pre-defined threshold (e.g., 15\%).
\item
{\em Selection}, which selects one or multiple sealed segments for GC.  In
this work, we focus on two selection algorithms: (i) Greedy
\cite{rosenblum92}, which selects the sealed segments with the highest GPs,
and (ii) Cost-Benefit \cite{rosenblum92,rumble14}, which selects the sealed
segments that have the highest values $\tfrac{GP * age}{1-GP}$ (where $age$
refers to the elapsed time of a sealed segment since it is sealed) for GC. 
\item
{\em Rewriting}, which discards all invalid blocks from the selected sealed
segments and writes back the remaining valid blocks into one or multiple open
segments.  The space of the selected sealed segments can then be reused. 
\end{itemize}

A log-structured storage system sees two types of written blocks:
each request that writes or updates an LBA in the workload generates one {\em
user-written block} (i.e., a new block) and zero or more {\em GC-rewritten
blocks} that are due to the rewrites of the block during GC. Thus, GC incurs
{\em write amplification (WA)}, defined as the ratio of the total number of
both user-written blocks and GC-rewritten blocks to the number of
user-written blocks.  In the deployment at Alibaba Cloud ESSDs
(\S\ref{sec:intro}), we observe that the high WA from GC degrades both the
effective I/O bandwidth and the SSD lifespans.  It is thus critical to
minimize WA.

\begin{figure}[!t]
\centering
\includegraphics[width=3.3in]{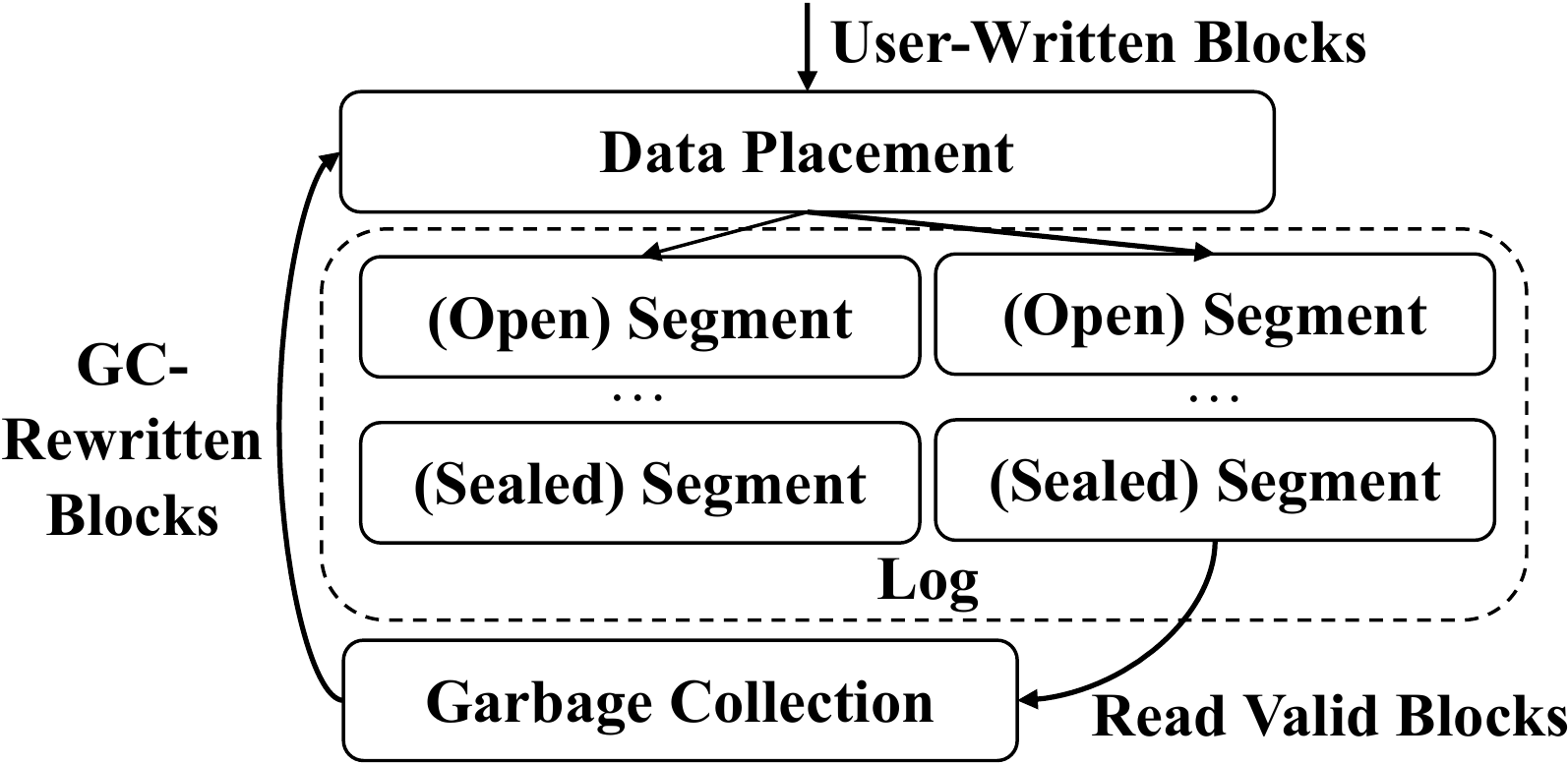}
\vspace{-3pt}
\caption{The workflow of a general data placement scheme.}
\label{fig:placement}
\vspace{-3pt}
\end{figure}

In this work, we aim to design a general and lightweight data
placement scheme that mitigates the WA due to GC in cloud-scale
deployment.
Figure~\ref{fig:placement} shows the workflow of a
general data placement scheme, which separates all written blocks (i.e.,
user-written blocks and GC-rewritten blocks) into different groups and writes
the blocks to the open segments of the respective groups.  The data placement
scheme is compatible with any GC policy (i.e., independent of the triggering,
selection, and rewriting policies).

\subsection{Ideal Data Placement}
\label{subsec:ideal}

We present an ideal data placement scheme that minimizes WA (i.e., WA$=$1).
We also elaborate why it is infeasible to realize in practice, so as to 
motivate the design of an effective practical data placement scheme.

\paragraph{System model.}  We first define the notations. Consider a write-only
request sequence of blocks that are written to a log-structured storage system.
Let $m$ be the number of user-written blocks
in the request sequence and $s$ be the
segment size (in units of blocks).  Let $k =
\lceil\tfrac{m}{s}\rceil$ be the number of sealed segments in the system, and
let $S_1, S_2, \cdots, S_k$ denote the corresponding $k$ sealed segments.  Let
$o_i$ (where $o_i\ge 1$) be the {\em invalidation order} of the $i$-th block in
the request sequence based on the BITs of all blocks (where $1\le i\le m$),
meaning that the $i$-th block is the $o_i$-th invalidated block among all
invalid blocks. 

\begin{figure}[t]
\centering
\includegraphics[width=3.3in]{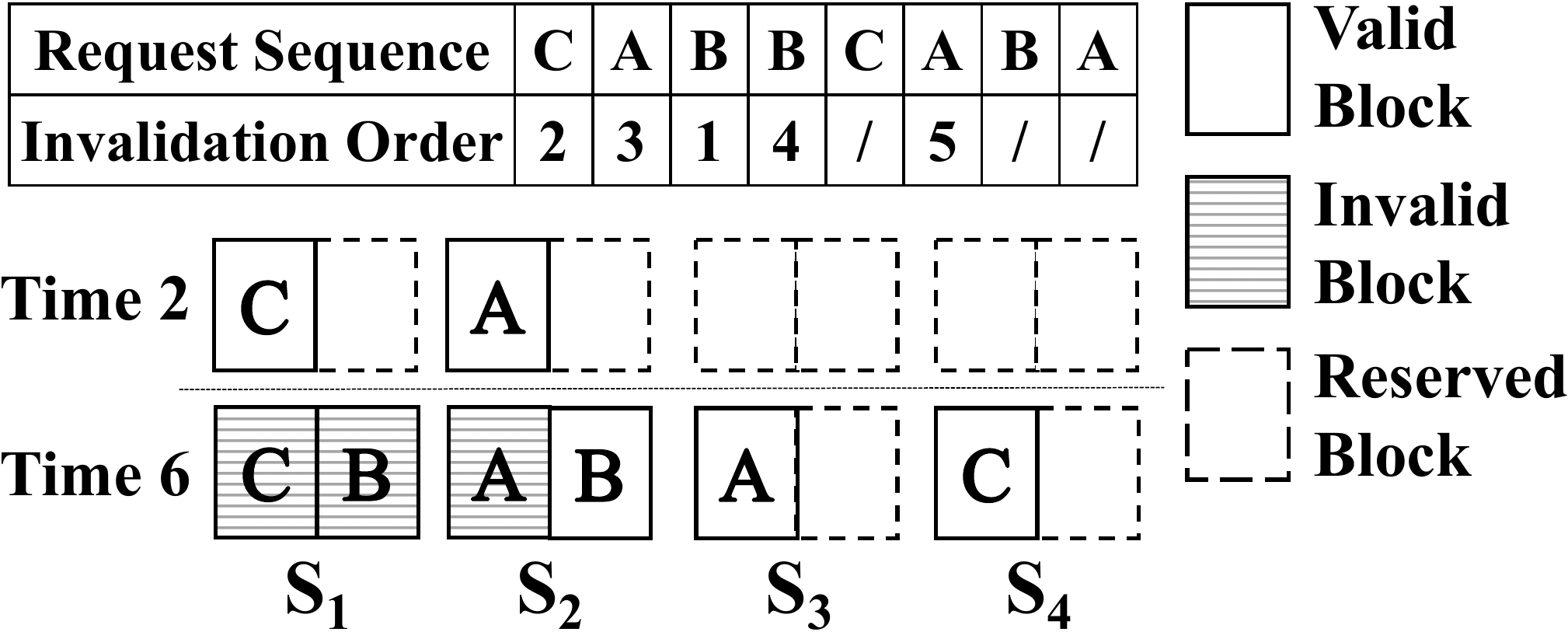}
\vspace{-3pt}
\caption{Example of the ideal data placement scheme.}
\label{fig:ideal}
\end{figure}

\begin{figure*}[!t]
\begin{tabular}{@{\ }ccc}
\parbox[t]{2.2in}{
\centering
\includegraphics[width=2.15in]{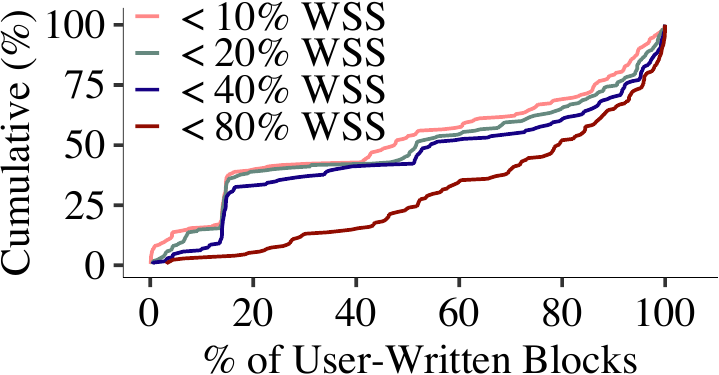}
\vspace{-3pt}
\caption{Percentages of user-written blocks with different short lifespans.}
\label{fig:motivation_ud}
} 
& 
\parbox[t]{2.2in}{
\centering
\includegraphics[width=2.15in]{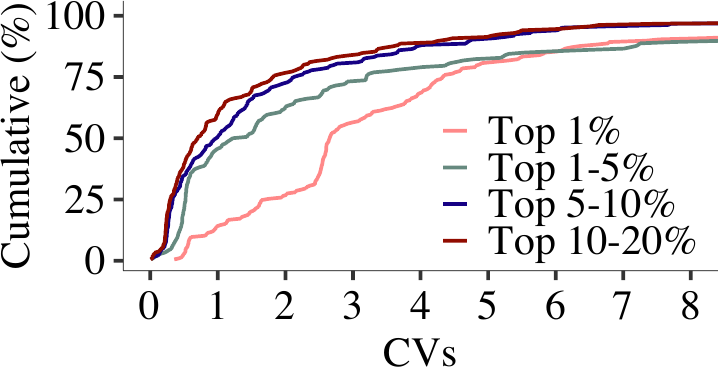}
\vspace{-3pt}
\caption{CVs of the lifespans of frequently updated blocks.}
\label{fig:motivation_cv}
} 
&
\parbox[t]{2.2in}{
\centering
\includegraphics[width=2.15in]{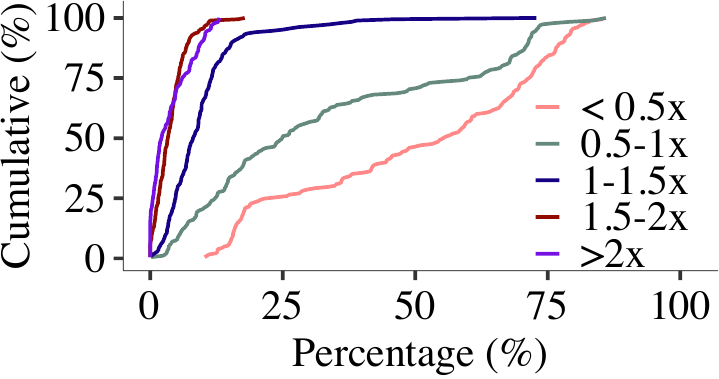} 
\vspace{-3pt}
\caption{Percentages of rarely updated blocks with different lifespans.}
\label{fig:motivation_ud_cold}
} 
\end{tabular}
\end{figure*}

\paragraph{Placement design.}  For the ideal placement scheme, we make the
following assumptions.  Suppose that the system has the future knowledge of the
BITs of all blocks, and hence the invalidation order $o_i$ of
the $i$-th block in the request sequence (where $1\le i\le m$). It also
allocates $k$ open segments for storing incoming blocks, and performs a GC
operation whenever there are $s$ invalid blocks in the system (i.e., one
segment size of invalid blocks). 

The system writes the $i$-th block to the $\lceil\tfrac{o_i}{s}\rceil$-th open
segment. If the $j$-th (where $1\le j\le k$) open segment is full, it is sealed
into the sealed segment $S_j$.  Thus, $S_j$ stores the blocks with the
invalidation orders in the range of $[(j-1)\cdot s + 1, j\cdot s]$.  The first
GC operation is triggered when there exist $s$ invalid blocks; according to the
placement, all such blocks must be stored in $S_1$.  Thus, the first GC
operation will choose $S_1$ for GC, and there will be no rewrites as all blocks
in $S_1$ must be invalid.  In general, the $j$-th GC operation (where $1\le
j\le k$) will choose $S_j$ for GC, and there will be no rewrites as $S_j$
contains only invalid blocks. 

Figure~\ref{fig:ideal} depicts an example of the ideal data placement scheme.
Consider a write-only request sequence with $m=8$ blocks with three LBAs $A$,
$B$, and $C$, and the $i$-th block is written at time~$i$ (where 
$1\le i\le m$). We fix the segment size as $s=2$. We show the status of the
volume at time~2 and time~6 when the second block and the sixth block are
written, respectively.  At time~2, we have appended $C$ to $S_1$ and $A$ to
$S_2$, as their invalidation orders are 2 and 3, respectively. Note that all
blocks in $S_1$ become invalid when block~C is updated at time~5,
and at this time we can perform a GC operation to reclaim the free space
occupied by $S_1$.  Note that the GC operation does not incur any rewrite.
Later, at time~6, the system appends $A$ to $S_3$ since its invalidation order
is 5.

\paragraph{Limitations and lessons learned.}  While the ideal data placement
scheme achieves the minimum WA, there exist two practical
limitations.  First, the scheme needs to have future knowledge of the BIT of
every block to assign the blocks to the corresponding open segments, but
having such future knowledge is infeasible in practice.  Second, 
the scheme needs to provision $k = \lceil m / s\rceil$ open segments to hold
all $m$ blocks in the request sequence in the worst case, as well as $k$
corresponding sealed segments for keeping the blocks from the $k$ open
segments.  Such provisioning incurring high memory and storage costs as $m$
increases.  Also, having too many open and sealed segments incurs substantial
random writes that lead to performance slowdown.

A practical data placement scheme should address the above two limitations.
Without the future knowledge of BITs, it should effectively {\em infer}
the BIT of each written block. With only a limited number of available open
segments, it should group written blocks by {\em similar BITs} instead of
placing them in strict invalidation order.  Our goal is to address the
limitations driven by real-world cloud block storage workloads.

\subsection{Trace Overview}
\label{subsec:traces}

We consider the {\em public} block-level I/O traces from two cloud block storage
systems, Alibaba Cloud \cite{li20} and Tencent Cloud
\cite{zhang20osca}.  The Alibaba Cloud traces
contain I/O requests (in multiples of 4\,KiB blocks) from 1,000 virtual disks,
referred to as {\em volumes}, over a one-month period in January 2020.
The Tencent Cloud traces have 4,995 volumes over a nine-day
period in October 2018. In this paper, we mainly focus on the Alibaba Cloud
traces, while we verified that the Tencent Cloud traces show similar findings 
(Appendix~\ref{sec:app_tencent}). 

The Alibaba Cloud traces comprise a variety of workloads (e.g., virtual
desktops, web services, key-value stores, and relational databases), and hence
are representative to drive our analysis. We treat each volume in the traces as
a standalone volume in the log-structured storage system (\S\ref{subsec:gc}),
such that each volume performs data placement and GC independently. Our goal is
to mitigate the overall WA across all volumes.

We pre-process the traces for our analysis and evaluation as follows.  We only
consider write requests as they are the only contributors of WA.  Since some
volumes in the traces have limited write requests to trigger sufficient GC
operations, we remove such volumes to avoid biasing our analysis.
Specifically, we focus on the volumes with sufficient write requests: each
volume has a write working set size (WSS) (i.e., the number of unique LBAs
being written multiplied by the block size) above 10\,GiB and a total write
traffic size (i.e., the number of written bytes) above $2\times$ its write
WSS. To this end, we select 186 volumes from the Alibaba Cloud
traces, which account for a total of over 90\% of write traffic of all 1,000
volumes.  The 186 volumes contain 10.9~billion write requests, 410.2\,TiB of
written data (with 390.2\,TiB of updates), 20.3\,TiB of write WSS (with
17.2\,TiB of update WSS).  Each of the 186 volumes has a write WSS ranging from
10\,GiB to 1\,TiB and a write traffic size ranging from 43\,GiB to 36.2\,TiB.
Since the WSS varies across volumes, we configure the maximum storage space of
each volume as $\frac{WSS}{1-GPT}$, where GPT denotes the GP threshold to
trigger GC.

\subsection{Motivation}
\label{subsec:motivation}

We show via trace analysis that existing data placement schemes cannot
accurately capture the BIT pattern and group the blocks with similar BITs for
effective WA mitigation.  We consider the 186 selected volumes from the Alibaba
Cloud traces (\S\ref{subsec:traces}). 
We define the lifespan of a block as the number of bytes written by the
workload from when a block is written until it is invalidated (or until the
end of the trace). A block is invalidated when the workload updates the
same LBA. We make three key observations.

\paragraph{Observation~1: User-written blocks generally have short lifespans.}
We say that a block has a {\em short lifespan} if its lifespan is smaller than
the write working set size (WSS) (i.e., the number of unique written
LBAs multiplied by the 4\,KiB block size).  We examine the percentages of
user-written blocks that fall into different lifespan range groups with short
lifespans that are represented as the fractions of the write WSS
for each volume.
Figure~\ref{fig:motivation_ud} shows the cumulative distributions
of the percentages of user-written blocks across all volumes in different
lifespan groups. In a large fraction of volumes, their user-written blocks
tend to have short lifespans.  For example, half of the volumes have more than
79.5\% of user-written blocks with lifespans smaller than
80\% of their write WSSes, and have more than
47.6\% of user-written blocks with lifespans smaller than
only 10\% write WSS. 
In contrast, GC-rewritten blocks generally have long lifespans. By definition,
GC-rewritten blocks are rewritten as they remain valid in the
GC-reclaimed segments. In both Greedy and Cost-Benefit selection algorithms,
GC tends to select segments that either show a high GP or exist for a long
time, implying that GC-rewritten blocks tend to have long lifespans.

Our findings suggest that user-written blocks and GC-rewritten blocks can have
vastly different BIT patterns, in which user-written blocks tend to have short
lifespans, while GC-rewritten blocks tend to have long lifespans.  Existing
data placement schemes either mix user writes and GC writes \cite{chiang99,
min12, stoica13, kremer19}, or focus on user writes \cite{shafaei16,
yang17, yang19}, in the data placement decisions.  Failing to distinguish
between user-written blocks and GC-rewritten blocks can lead
to inefficient WA mitigation. Instead, it is critical to separately identify
the BIT patterns of user-written blocks and GC-rewritten blocks. 

\paragraph{Observation~2: Frequently updated blocks have highly varying
lifespans.} We investigate {\em frequently updated blocks}, referred to as the
blocks whose {\em update frequencies} (i.e., the number of updates) rank in the top
20\% in the write working set (i.e., the set of LBAs being written) of
a volume.  Specifically, for each volume, we divide the frequently updated
blocks into four groups based on their ranks of update frequencies, 
namely top 1\%, top 1-5\%, top 5-10\%, and top 10-20\%, so that the blocks in
each group have similar update frequencies.
The medians of the minimum update frequency in
the four groups across all volumes are 37.5, 8.5, 6.0, and 5.0, respectively. To
avoid evaluation bias, we exclude the blocks that have not been invalidated
before the end of the traces.  For each group of a volume, we calculate the
{\em coefficient of variation (CV)} (i.e., the standard deviation divided by the
mean) of the lifespans of the blocks; a high CV (e.g., larger than one) implies
a high variance in the lifespans.

Figure~\ref{fig:motivation_cv} shows the cumulative distributions
of CVs across all volumes (note that 6, 6, 20, and 18 volumes in the four
groups have CVs exceeding 8, respectively). We see that frequently
updated blocks with similar update frequencies have high variance in their
lifespans (and hence the BITs); for example, 25\% of the volumes have their CVs
exceeding 4.34, 3.20, 2.14, and 1.82 in the four groups top 1\%, top 1-5\%, top
5-10\%, and top 10-20\%, respectively. Our findings also suggest that existing
temperature-based data placement schemes that group the blocks with similar
write/update frequencies
\cite{chiang99,min12,stoica13,shafaei16,yang17,kremer19,yang19} cannot
effectively group blocks with similar BITs, and hence the WA cannot be fully
mitigated. 

\paragraph{Observation~3: Rarely updated blocks dominate and have highly
varying lifespans.}
We examine the write working set of each volume and define the {\em rarely
updated blocks} as those that are updated no more than four times during the
one-month trace period. We see that rarely updated blocks occupy a high
percentage in the write working sets of a large fraction of volumes.  In
half of the volumes, more than 72.4\% of their write working sets contain
rarely updated blocks.  We further examine the lifespans of those rarely
updated blocks.  For each volume, we divide the rarely updated blocks into
five groups that are partitioned by the lifespans of 0.5$\times$, 1$\times$,
1.5$\times$, and 2$\times$ of their write WSSes.  We then calculate the
percentage of those blocks that fall into each group.

Figure~\ref{fig:motivation_ud_cold} shows the cumulative distributions of the
percentages of rarely updated blocks in different lifespan groups across all
volumes.  In 25\% of the volumes, more than 71.5\% of the rarely updated
blocks have their lifespans smaller than 0.5$\times$ write WSS. For the
remaining four groups, the medians of the percentages are 24.9\%, 8.1\%,
3.3\%, and 2.2\%, respectively.  In other words, the lifespans of rarely
updated blocks can span both short and long lifespan ranges, and hence
show high deviations of BITs in a volume.
As in Observation~2, our findings again suggest that existing temperature-based
data placement schemes cannot effectively group the rarely updated blocks with
similar BITs.  Rarely updated blocks are often treated as cold blocks with low
write frequencies, so they tend to be grouped together and separated from the
hot blocks with high write frequencies.  However, their vast differences in BIT
patterns make temperature-based data placement schemes inefficient in
mitigating WA.

\section{\sysname Design}
\label{sec:design}

\subsection{Design Overview}
\label{subsec:overview}

\begin{figure}[t]
\centering
\includegraphics[width=3.3in]{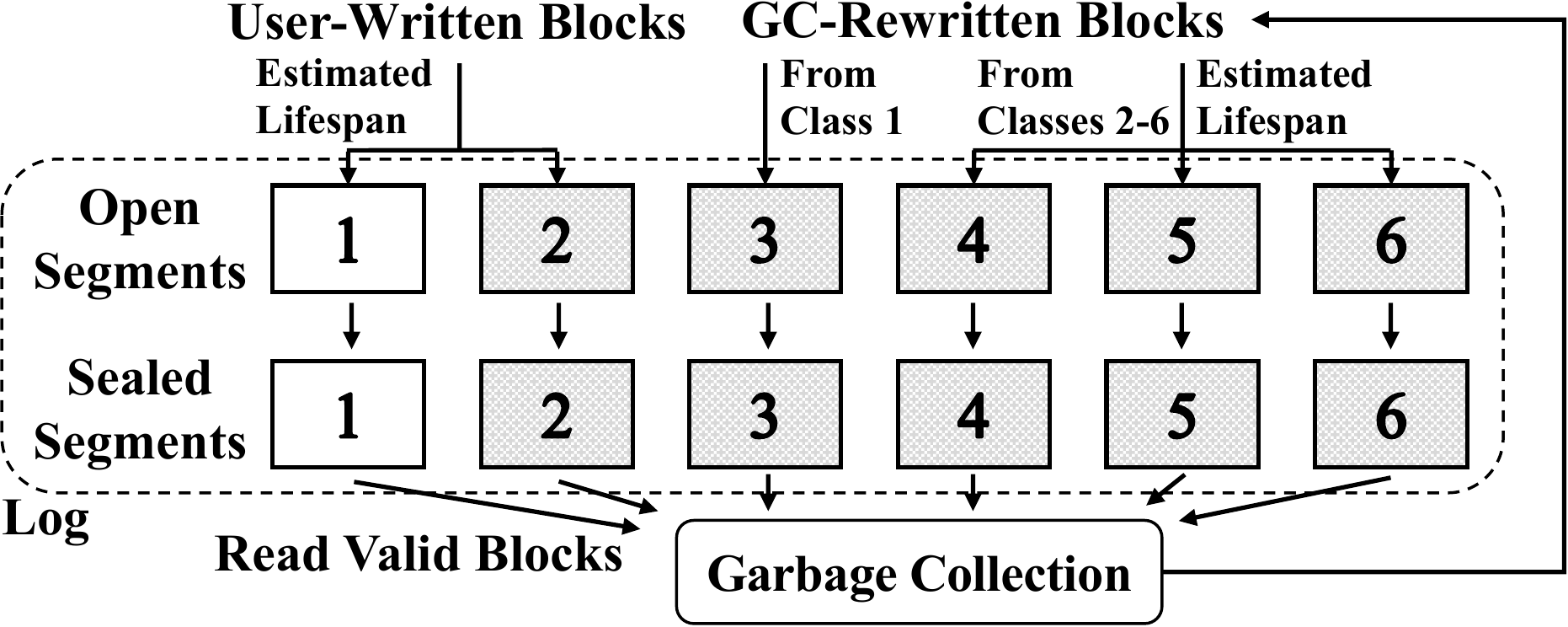} 
\vspace{-3pt}
\caption{\sysname workflow.}
\label{fig:sepbit_flow}
\vspace{-10pt}
\end{figure}

We design \sysname based on our observations in
\S\ref{subsec:motivation}.  \sysname first separates blocks into user-written
blocks and GC-rewritten blocks due to their different BIT patterns
(Observation~1).  It further separates both user-written blocks and
GC-rewritten blocks by inferring their BITs instead of using block
temperatures as in existing temperature-based approaches (Observations~2 and
3).

Figure~\ref{fig:sepbit_flow} depicts the workflow of \sysname.  Our current
design of \sysname defines {\em six} classes of segments, in which Classes~1-2
correspond to the segments of user-written blocks, while Classes~3-6
correspond to the segments of GC-rewritten blocks. Each class is now
configured with one open segment and has multiple sealed segments.  If an open
segment reaches the maximum size, it is sealed and remains in the same class.

\sysname infers the lifespans of blocks and in turn the corresponding BITs of
blocks.  For user-written blocks (i.e., Classes~1-2), \sysname stores the {\em
short-lived} blocks (with short lifespans) in Class~1 and the remaining {\em
long-lived} blocks (with long lifespans) in Class~2.  For GC-rewritten blocks
(i.e., Classes~3-6), \sysname appends the blocks from Class~1 that are
rewritten by GC into Class~3, and groups the remaining GC-rewritten blocks
into Classes~4-6 by similar BITs inferred.  

\begin{figure}[t]
\centering
\begin{tabular}{c}
\includegraphics[width=3.2in]{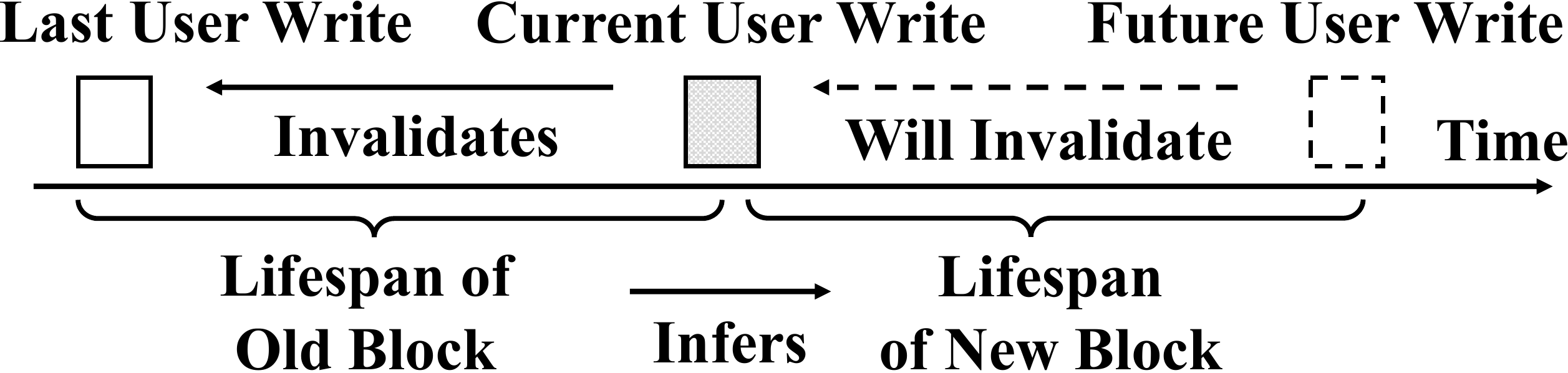}\\
\mbox{\small (a) Inferring BITs of user-written blocks}
\vspace{6pt}\\
\includegraphics[width=3.2in]{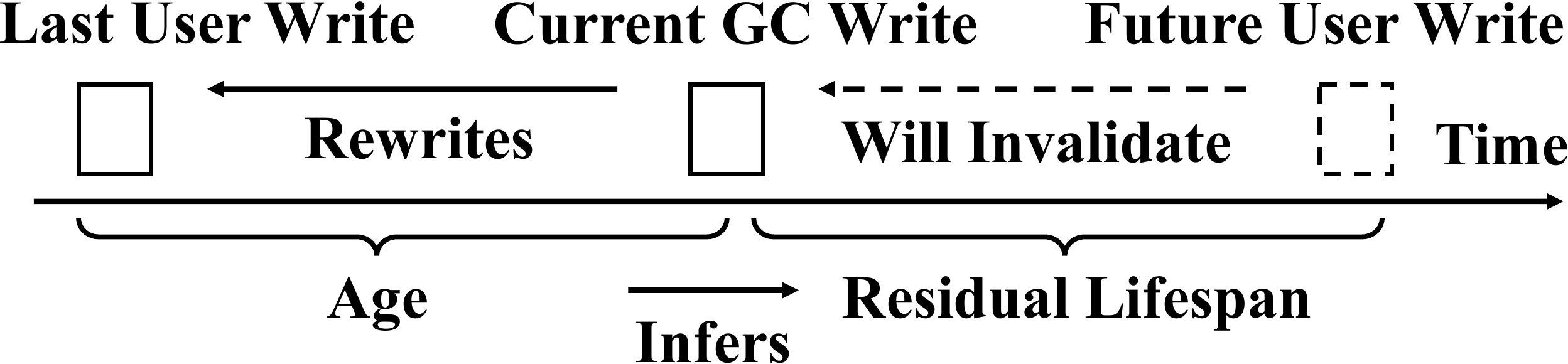}\\
\mbox{\small (b) Inferring BITs of GC-rewritten blocks}
\end{tabular}
\vspace{-8pt}
\caption{Ideas of inferring BITs in \sysname.}
\vspace{-10pt}
\label{fig:infer}
\end{figure}

The main idea of \sysname is as follows.  For each user-written
block, \sysname examines its {\em last user write time} to infer its lifespan.
Specifically, for the write time, \sysname uses a monotonic timer
(instead of the real timestamp) that increments by one for each user-written
block.
If the user-written block is
issued from a new write, \sysname assumes that it has an infinite lifespan.
Otherwise, if the user-written block updates an old block, \sysname uses the
lifespan of the old block (i.e., the number of user-written bytes in the whole
workload since its last user write time until it is now invalidated) to
estimate the lifespan of the user-written block, as shown in
Figure~\ref{fig:infer}(a).  Our intuition is that {\em any user-written block
that invalidates a short-lived block is also likely to be a short-lived block}
(\S\ref{subsec:userwrite}). Then if the short-lived blocks are written at about
the same time, their corresponding BITs will be close, so \sysname groups them
into same class (i.e., Class~1).  For the long-lived blocks (including the
user-written blocks from new writes), \sysname groups them into Class~2.

For each GC-rewritten block, \sysname examines its {\em age}, defined as the
number of user-written bytes in the whole workload since its
last user write time until it is rewritten by GC, to infer its {\em residual
lifespan}, defined as the number of user-written bytes since it
is rewritten by GC until it is invalidated (or until the end of the traces),
as shown in Figure~\ref{fig:infer}(b). As a result, the lifespan of a
GC-rewritten block is its age plus its residual lifespan. Our intuition is
that {\em any GC-rewritten block with a smaller age has a higher probability
to have a short residual lifespan} (\S\ref{subsec:gcwrite}), implying that
GC-rewritten blocks with different ages are expected to have different
residual lifespans. Thus, \sysname can distinguish the blocks of different
residual lifespans based on their ages and group the GC-rewritten blocks with
similar ages into the same classes.

Our design builds on the assumption that the access pattern is {\em skewed}
for inferring the BITs of blocks. We justify our assumption via the
mathematical analysis for skewed distributions and the trace analysis for
real-world workloads (\S\ref{subsec:userwrite} and \S\ref{subsec:gcwrite}).
To adapt to changing workloads and GC policies, \sysname monitors the
workloads to separate user-written blocks and GC-rewritten blocks into
different classes (\S\ref{subsec:impl}).

\subsection{Inferring BITs of User-Written Blocks}
\label{subsec:userwrite}

We show via both mathematical and trace analyses the effectiveness of \sysname
in estimating the BITs of user-written blocks based on the lifespans.  Let $n$
be the total number of unique LBAs in a working set; without loss of
generality, each LBA is denoted by an integer from 1 to $n$.  Let $p_i$ (where
$1\le i\le n$) be the probability that LBA~$i$ is being written 
in each write request.  Consider a write-only request sequence of blocks, each
of which is associated with a sequence number $b$ and the LBA $A_{b}$.  Let
$b$ and $b'$ (where $b' < b$) denote the sequence numbers of a new
user-written block and the corresponding invalid old block, respectively
(i.e., $A_{b}=A_{b'}$).

Recall from \S\ref{subsec:overview} that \sysname estimates the lifespan
(denoted by $u$) of the user-written block $b$ using the lifespan (denoted by
$v$) of the old block $b'$, so the estimated BIT of block $b$ is equal to the
current user write time plus the estimated lifespan $u$; note that both $u$
and $v$ are expressed in units of blocks.  We claim that if $v$ is small, $u$
is also likely to be small.  To validate the claim, let $u_0$ and $v_0$ (both
in units of blocks) be two thresholds.  We then examine the conditional
probability of $u \le u_0$ given the condition that $v \le v_0$ subject to a
workload of different skewness.  If the conditional probability is high for
small $u_0$ and $v_0$, then our claim holds. 

\paragraph{Mathematical analysis.} We examine the following conditional
probability (see derivation in Appendix~\ref{sec:app_derivation}):
\begin{align*}
  &\Pr(u \le u_0 \mid v \le v_0) = \frac{\Pr(u \le u_0 \text{ and } v \le v_0)} {\Pr(v \le v_0)} \\
  &= \frac
     {\sum\nolimits_{i=1}^{n} (1 - (1 - p_i)^{u_0}) \cdot (1 - (1 - p_i)^{v_0}) \cdot p_i}
     {\sum\nolimits_{i=1}^{n} (1 - (1 - p_i)^{v_0}) \cdot p_i}.
\end{align*}

We analyze the conditional probability via the Zipf distribution, given by
$p_i=(1/i^{\alpha}) / \sum_{j=1}^{n}(1/j^{\alpha})$, where $1 \le i \le n$ for
some skewness parameter $\alpha \ge 0$. A larger $\alpha$ implies a more
skewed distribution.  We fix $n=10 \times 2^{18}$, which corresponds to a
working set of 10\,GiB with 4\,KiB blocks. We then study how the conditional
probability $\Pr(u \le u_0 \mid v \le v_0)$ varies across $u_0$, $v_0$, and
$\alpha$.
 
\begin{figure}[t]
\centering
\begin{tabular}{@{\ }cc}
\includegraphics[width=1.4in]{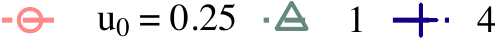}
&
\includegraphics[width=1.4in]{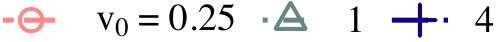}\\
\includegraphics[width=1.6in, valign=t]{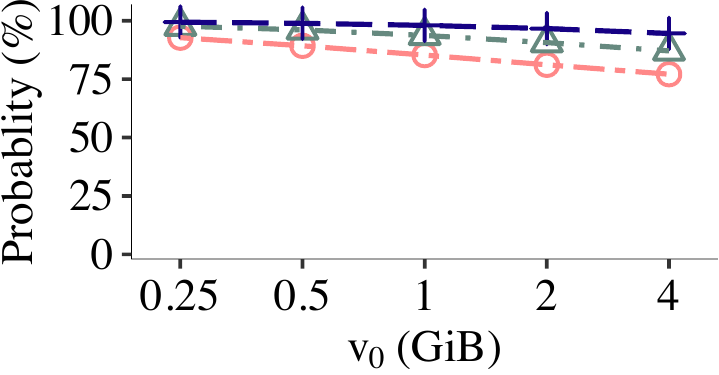} &
\includegraphics[width=1.6in, valign=t]{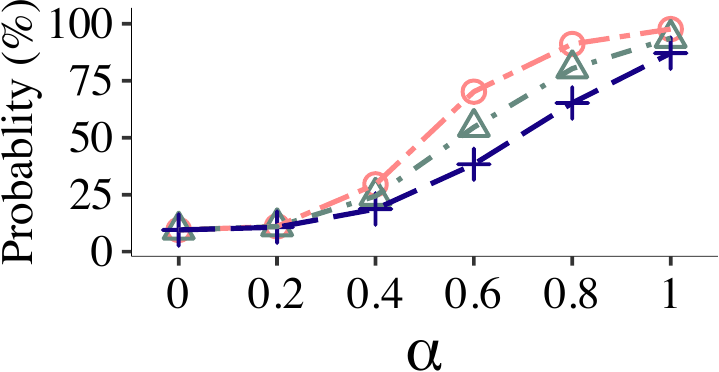} \\
\mbox{\small (a) $\alpha=1$; varying $u_0$, $v_0$} &
\mbox{\small (b) $u_0=1$\,GiB; varying $v_0$, $\alpha$}
\end{tabular}
\vspace{-6pt}
\caption{Inferring BITs of user-written blocks: $\Pr(u \le u_0 \mid v \le
v_0)$ versus $v_0$ and $\alpha$.}
\label{fig:userwrite_vary}
\end{figure}

Figure~\ref{fig:userwrite_vary}(a) first shows the conditional probability for
varying $u_0$ and $v_0$, where we fix $\alpha=1$.  We focus on short lifespans
by varying $u_0$ and $v_0$ of up to 4\,GiB, which is less than the write WSS
(\S\ref{subsec:motivation}).  Overall, the conditional probability is high for
different $u_0$ and $v_0$; the lowest one is 77.1\% for $v_0 = 4$\,GiB and
$u_0 = 0.25$\,GiB.  This shows that a user-written block is highly likely to 
have a short lifespan if its invalidated block also has a short lifespan. 
In particular, the conditional probability is higher if $v_0$ is smaller
(i.e., the invalidated blocks have shorter lifespans), implying a more
accurate estimation of the lifespan of the user-written block.

Figure~\ref{fig:userwrite_vary}(b) next shows the conditional probability for
varying $v_0$ and $\alpha$, where we fix $u_0=1$\,GiB.  Note that for $\alpha
= 0$, the Zipf distribution reduces to a uniform distribution.  Overall, the
conditional probability increases with $\alpha$ (i.e., more skewed).  For
example, for $\alpha = 1$, the conditional probability is at least 87.1\%.
However, for $\alpha=0$, the conditional probability is only 9.5\%.  This
indicates that the high accuracy of lifespan estimation only holds under
skewed workloads.

\begin{figure}[t]
\centering
\includegraphics[width=3.2in]{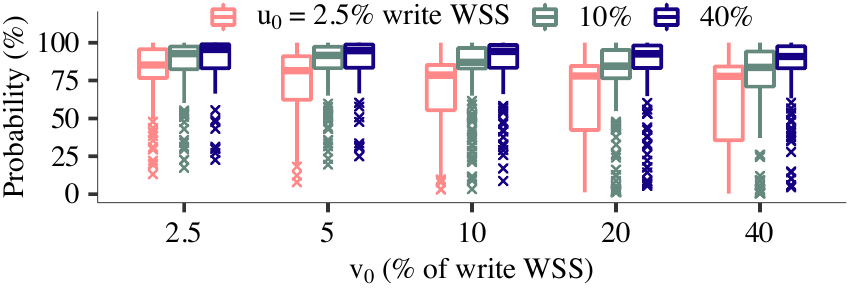}
\vspace{-6pt}
\caption{Inferring BITs of user-written blocks: Boxplots of 
$\Pr(u \le u_0 \mid v \le v_0)$ for different $u_0$ and $v_0$ in real-world
workloads.}
\label{fig:trace_userwrite}
\vspace{-6pt}
\end{figure}

\paragraph{Trace analysis.}  We use the block-level I/O traces from Alibaba
Cloud (\S\ref{subsec:traces}) to validate if the conditional probability
remains high in real-world workloads.  To compute the conditional probability, 
we first find the set of user-written blocks that invalidate old blocks
with $v \le v_0$.  Then the conditional probability is the fraction of blocks
with $u \le u_0$ in the set. We vary both $v_0$ and $u_0$ as
different percentages of the write WSS to examine different conditional
probabilities.
Figure~\ref{fig:trace_userwrite} shows the boxplots of the conditional
probabilities over all volumes for different $u_0$ and $v_0$.  In general, the
conditional probability remains high in most of the volumes. For example, for
$v_0$ being 40\% of write WSS, the medians of the conditional
probabilities are in the range of 77.8-90.9\%, and the 75th percentiles are in
the range of 84.3-97.6\%. Also, the conditional probability tends to be
higher for a smaller $v_0$. 

\subsection{Inferring BITs of GC-Rewritten Blocks}
\label{subsec:gcwrite}

We further show via both mathematical and trace analyses the effectiveness of
\sysname in estimating the BITs of GC-rewritten blocks based on the {\em
residual} lifespans. Recall from \S\ref{subsec:overview} that \sysname
estimates the residual lifespan of a GC-rewritten block using its age, so the
estimated BIT of the GC-rewritten block is equal to the current GC write time
plus the estimated residual lifespan.  However, characterizing directly
GC-rewritten blocks is non-trivial, as it depends on the actual GC policy
(e.g., when GC is triggered and which segments are selected for GC)
(\S\ref{subsec:gc}).  Instead, we model GC-rewritten blocks based on 
user-written blocks.  If a user-written block has a lifespan above a certain
threshold, we assume that it is rewritten by GC and treat it as a GC-rewritten
block with an age equal to the threshold.  We can then apply a similar
analysis for user-written blocks as in \S\ref{subsec:userwrite}.

We define the following notations. As each GC-rewritten block is a
user-written block before being rewritten by GC, we identify each GC-rewritten
block by its corresponding user-written block with sequence number $b$. Let
$u$, $g$, and $r$ be its lifespan, age, and residual lifespan,
respectively, such that $u = g + r$; each of the variables is measured in units
of blocks.  We claim that $r$ has a higher probability to be small with a
smaller $g$. To validate the claim, let $g_0$ and $r_0$ (both in units of
blocks) be the thresholds for the age and the residual lifespan, respectively.
We examine the conditional probability of $u \le g_0 + r_0$ given the condition
that $u \ge g = g_0$ subject to a workload of different skewness. The
conditional probability specifies the fraction of GC-rewritten blocks whose
residual lifespans are shorter than $r_0$ among all GC-rewritten blocks with
age $g_0$ (note that the GC-rewritten blocks are modeled as user-written blocks
with lifespans above $g_0$).  If the conditional probability is higher for a
smaller $g_0$ subject to a fixed $r_0$, then our claim holds.

\paragraph{Mathematical analysis.} We examine the following conditional
probability (see derivation in Appendix~\ref{sec:app_derivation}):
\begin{align*}
  & \Pr(u \le g_0 + r_0 \mid u \ge g_0) = \frac{\Pr(g_0 \le u \le g_0 + r_0)}{\Pr(u \ge g_0)} \\
  & = \frac{\sum\nolimits_{i=1}^{n} p_i \cdot ((1-p_i)^{g_0} - (1-p_i)^{g_0 + r_0})}
      {\sum\nolimits_{i=1}^{n} p_i \cdot (1-p_i)^{g_0}}.
\end{align*}

As in \S\ref{subsec:userwrite}, we use the Zipf distribution and fix $n =
10 \times 2^{18}$ unique LBAs. We study how the conditional probability $\Pr(u \le g_0 +
r_0 \mid u \ge g_0)$ varies across $g_0$, $r_0$, and $\alpha$.

\begin{figure}[t]
\centering
\begin{tabular}{@{\ }cc}
\includegraphics[width=1.3in]{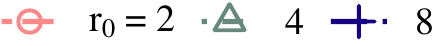}
&
\includegraphics[width=1.3in]{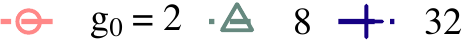}\\
\includegraphics[width=1.6in]{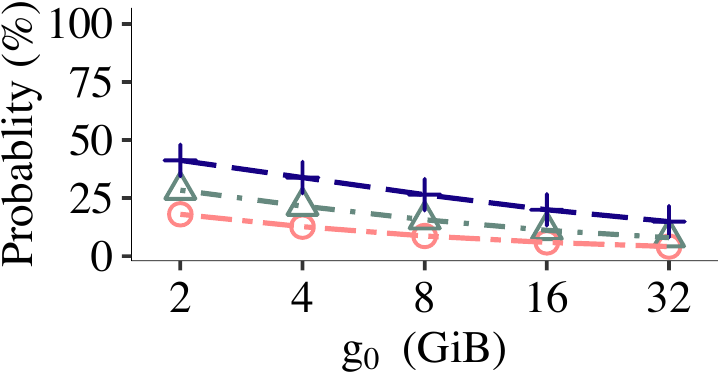} &
\includegraphics[width=1.6in]{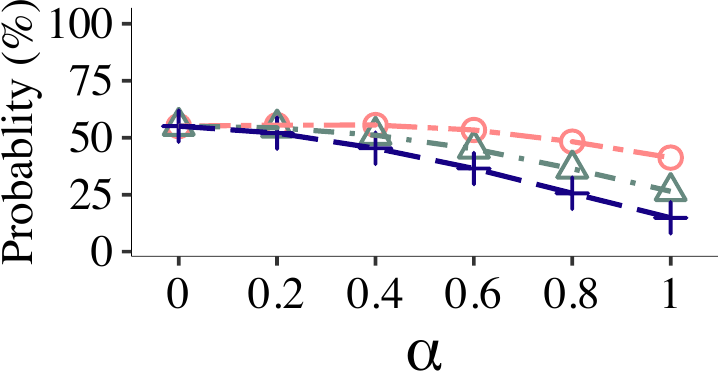}\\
\mbox{\small (a) $\alpha=1$; varying $r_0$, $g_0$} &
\mbox{\small (b) $r_0=8$\,GiB; varying $g_0$, $\alpha$}
\end{tabular}
\vspace{-6pt}
\caption{Inferring BITs of GC-rewritten blocks: 
$\Pr(u \le g_0 + r_0 \mid u \ge g_0)$ versus $g_0$ and $\alpha$.}
\label{fig:gcwrite_vary}
\end{figure}

Figure~\ref{fig:gcwrite_vary}(a) first shows the conditional probability for
varying $g_0$ and $r_0$, where we fix $\alpha = 1$.  We focus on a large value
of $g_0$ of up to 32\,GiB since we target long-lived blocks.  We also vary
$r_0$ up to 8\,GiB. Overall, for a fixed $r_0$, the conditional probability
decreases as $g_0$ increases.  For example, given that $r_0 = 8$\,GiB, the
probability with $g_0 = 2$\,GiB is 41.2\%, while the probability for $g_0 =
32$\,GiB drops to 14.9\%.  This validates our claim that GC-rewritten blocks
with different ages are expected to have different residual lifespans.  Thus,
we can distinguish the GC-rewritten blocks of different residual lifespans
based on their ages.

Figure~\ref{fig:gcwrite_vary}(b) further shows the conditional probability for
varying $g_0$ and $\alpha$, where we fix $r_0=8$\,GiB.  For a small $\alpha$,
the conditional probability has a limited difference for varying $v$, while the
difference becomes more significant as $\alpha$ increases.  For example, for
$\alpha = 0$ (i.e., the uniform distribution), there is no difference varying $g_0$;
for $\alpha = 0.2$, the difference of the conditional probability between $g_0
= 2$\,GiB and $g_0 = 32$\,GiB is only 3.5\%, while the difference for $\alpha =
1$ is 26.4\%. This indicates that our claim holds under skewed workloads, and
we can better distinguish the GC-rewritten blocks of different residual
lifespans under more skewed workloads.

\begin{figure}[!t]
\centering
\includegraphics[width=3.2in]{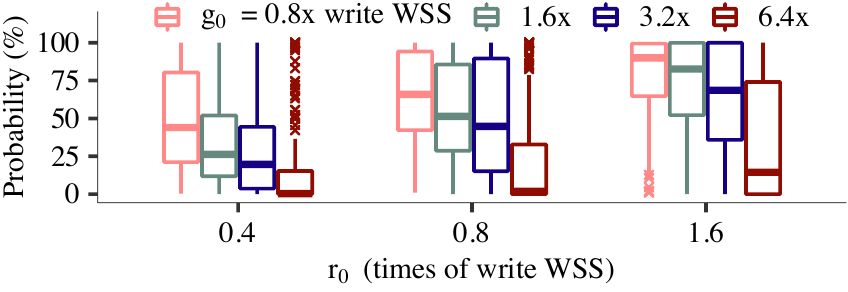}
\vspace{-6pt}
\caption{Inferring BITs of GC-rewritten blocks: Boxplots of 
$\Pr(u \le g_0 + r_0 \mid u \ge g_0)$
for different $r_0$ and $g_0$ in real-world workloads.}
\vspace{-8pt}
\label{fig:trace_gcwrite}
\end{figure}

\paragraph{Trace analysis.} We also use block-level I/O traces from Alibaba
Cloud (\S\ref{subsec:traces}) to examine the conditional probability in
real-world workloads.  We first identify the set of blocks with $u \ge g_0$ in
the workload, and then compute the conditional probability as a fraction of
blocks with $u \le g_0 + r_0$ in the set. We vary both $r_0$ and
$g_0$ as different percentages of the write WSS.
Figure~\ref{fig:trace_gcwrite} depicts the boxplots of the conditional
probabilities over all volumes for different $g_0$ and $r_0$.  For a fixed
$r_0$, the conditional probabilities have significant differences for varying
$g_0$. For example, if we fix $r_0$ as $1.6\times$ of write WSS
and $g_0$ increases from $0.8\times$ to $6.4\times$ of write WSS, the median
probabilities drop from 90.0\% to 14.5\%.


\subsection{Implementation Details}
\label{subsec:impl}

\noindent
{\bf Threshold selection.}  We assign blocks into different classes by
their estimated BITs with multiple thresholds: for user-written blocks, we
define a {\em lifespan threshold} for separating short-lived blocks and
long-lived blocks; for GC-rewritten blocks, we need multiple {\em age
thresholds} to separate them by ages (\S\ref{subsec:overview}).  We configure
the thresholds via the {\em segment lifespan} of a segment, defined as the
number of user-written bytes in the workload since the segment
is created (i.e., the time when the first block is appended to the segment)
until it is reclaimed by GC. Specifically, we monitor the average segment
lifespan, denoted by $\ell$, among a fixed number of recently
reclaimed segments in Class~1. For each user-written block, if it
invalidates an old block with a lifespan less than $\ell$, we write it to
Class~1; otherwise, we write it to Class~2. For GC-rewritten blocks, we set
the age thresholds as multiples of $\ell$ (see below).

\begin{algorithm}[t]
\caption{\sysname}\label{alg:sepbit}
\begin{small}
\begin{algorithmic}[1]
\State $t = 0$;\ $\ell=+\infty$;\ $\ell_{tot} = 0$;\ $n_{c} = 0$, where $t$ is the global timestamp
\Function{\textup{\texttt{GarbageCollect}}}{ }
  \State Select a segment $S$ by selection algorithm
  \If {$S$ is from Class~1}
	\State $n_{c} = n_{c}+1$,\ $\ell_{tot} = \ell_{tot} + (t - S.creation\_time)$
  \EndIf
  \If {$n_{c} = 16$}
	\State $\ell = \ell_{tot} / n_{c}$;\ $n_{c} = 0$;\ $\ell_{tot} = 0$
  \EndIf
  \For{each valid block $b$ in $S$}
	\State \texttt{GCWrite}($b$)
  \EndFor
\EndFunction
\Function{\textup{\texttt{UserWrite}}}{$b$}
  \State Find lifespan $v$ of the invalidated block $b'$ due to $b$
  \If{$v < \ell$}
	\State Append $b$ to open segment of Class~1
  \Else
	\State Append $b$ to open segment of Class~2
  \EndIf
  \State $t = t+1$
\EndFunction

\Function{\textup{\texttt{GCWrite}}}{$b$}
  \If{$b$ is from Class~1}
	\State Append $b$ to open segment of Class~3
  \Else
  \State $g = t - b.last\_user\_write\_time$
	\State If $g \in [0, 4\ell)$, append $b$ to open segment of Class~4
	\State If $g \in [4\ell, 16\ell)$, append $b$ to open segment of Class~5 
	\State If $g \in [16\ell, +\infty)$, append $b$ to open segment of Class~6
  \EndIf

\EndFunction
\end{algorithmic}
\end{small}
\end{algorithm}

\paragraph{Algorithmic details.} Algorithm~\ref{alg:sepbit} shows the
pseudo-code of \sysname, which consists of three functions:
\texttt{GarbageCollect}, \texttt{UserWrite}, and \texttt{GCWrite}.  Each class
always corresponds to one open segment.  If an open segment is full, it
becomes a sealed segment, and \sysname creates a new open segment within the
same class.  \sysname initializes the average segment lifespan $\ell=+\infty$,
which is updated on-the-fly.  It also tracks a global timestamp $t$, which
records the sequence number of the current user-written block.

\texttt{GarbageCollect} is triggered by a GC operation according to the GC
policy (\S\ref{subsec:gc}). It
performs GC and monitors the runtime information of the reclaimed segments. It
selects a segment $S$ for GC based on the selection algorithm (e.g., Greedy or
Cost-Benefit (\S\ref{subsec:gc})). It sums up the lifespans of collected
segments from Class~1 as $\ell_{tot}$, and computes the average lifespan
$\ell=\ell_{tot}/n_c$ for every fixed number $n_c$ (e.g., $n_c=16$ in our
current implementation) of reclaimed segments. 

\texttt{UserWrite} processes each user-written block $b$.  It first computes
the lifespan $v$ of the invalidated old block $b'$. If $v$ is less than
$\ell$, \texttt{UserWrite} appends $b$ (which is treated as a short-lived
block) to the open segment of Class~1; otherwise, it appends $b$ (which is
treated as a long-lived block) to the open segment of Class~2. 

\texttt{GCWrite} processes each GC-rewritten block that corresponds to some
user-written block $b$.  If $b$ is originally stored in Class~1,
\texttt{GCWrite} appends $b$ to the open segment of Class~3; otherwise,
\texttt{GCWrite} appends $b$ to one of the open segments of Classes~4-6 based
on the age of $b$.  Currently, we configure the age thresholds as three
ranges, $[0, 4\ell)$, $[4\ell, 16\ell)$, and $[16\ell, +\infty)$, for
Classes~4-6, respectively, based on our evaluation findings. Nevertheless, we
have also experimented with different numbers of classes and thresholds 
(Appendix \ref{sec:app_eval}), and we observe only marginal differences in WA.

\paragraph{Memory usage.}  \sysname only stores the last user write time of
each block as the metadata alongside the block {\em on disk}, without
maintaining a mapping from every LBA to its last user write time in memory.
Putting metadata alongside a block is feasible, as SSDs typically associate a
small spare region (e.g., of size 64~bytes) with each flash page for storing
metadata.  Specifically, for user-written blocks, \sysname only needs to know
whether the lifespan of an invalidated block is shorter than a threshold.  It
thus suffices for \sysname to track only the recently written LBAs.
In our current implementation (written in C++), \sysname maintains
a first-in-first-out (FIFO) queue to record recently written LBAs.
It dynamically adjusts the queue length according to the value $\ell$.  If the
FIFO queue is full, each insert of an element will dequeue one element from
the queue.  If $\ell$ increases, the FIFO queue allows more inserts without
dequeueing any element; if $\ell$ decreases, the FIFO queue dequeues two
elements for each insert until the number of elements drops below $\ell$.  If
the LBA exists in the FIFO queue and its user write time is within the recent
$\ell$ user writes, \sysname writes it into Class~1.
To efficiently query the FIFO queue, \sysname creates a {\tt
std::map} structure in the C++ standard template library to record each unique
LBA in the FIFO queue and its latest queue position. When we enqueue the LBA
of a newly written block into the FIFO queue, we insert or update the LBA with
its current queue position in the {\tt std::map} structure; when we dequeue
an LBA from the FIFO queue, we remove the LBA from the {\tt std::map}
structure if its recorded queue position is equal to the dequeued one.

For GC-rewritten blocks, \sysname retrieves them during GC
and examines the user write time directly from the metadata,
so as to assign the GC-rewritten block to the corresponding
class without any memory overhead incurred. 

\paragraph{Prototype.} We prototype a log-structured block storage system that
realizes \sysname and existing data placement schemes.  We choose to deploy
our prototype on zoned storage \cite{zonedstorage}, whose append-only
interfaces favor log-structured storage deployment.  Specifically, our
prototype runs on an emulated zoned storage backend based on ZenFS
\cite{zenfs} (due to the lack of a real zoned storage device, we currently
emulate the zoned storage backend using Intel Optane Persistent Memory \cite{optane}).  Each
segment in the prototype is a one-to-one mapping to a {\em ZoneFile}, the basic
unit in the zoned storage backend in ZenFS.  Then ZenFS stores ZoneFiles in
different zones without incurring device-level GC.  For the metadata and the
FIFO queue in \sysname, the prototype stores them in separate files and accesses
them using \texttt{mmap} for memory efficiency; for other existing data
placement schemes, the prototype stores all metadata in memory.  When the
prototype triggers GC (at the system level), it reads only valid blocks from
storage and rewrites the blocks into different segments.  

The reasons of choosing emulated zoned storage based on ZenFS 
in our prototype
are three-fold.  First, zoned storage has a similar storage abstraction to
Pangu (\S\ref{sec:intro}), as both of them support append-only writes and
large-size append-only units (e.g., up to hundreds of MiB).  Second, emulated
zoned storage provides minimal external interference, making the performance
evaluation reproducible; in contrast, the performance of traditional SSDs can
be easily disturbed by device-level GC.  Finally, ZenFS is a lightweight
user-space zone-aware file system that readily supports zoned storage.

\section{Evaluation}
\label{sec:eval}

\subsection{Data Placement Schemes}
\label{subsec:schemes}

We compare \sysname with eight existing temperature-based data placement
schemes, namely Dynamic dAta Clustering (DAC) \cite{chiang99}, SFS
\cite{min12}, MultiLog (ML) \cite{stoica13}, extent-based identification (ETI)
\cite{shafaei16}, MultiQueue (MQ) \cite{yang17}, Sequentiality, Frequency, and
Recency (SFR) \cite{yang17}, Fading Average Data Classifier (FADaC)
\cite{kremer19}, and WARCIP \cite{yang19}.   Note that these existing schemes
are mainly designed for mitigating the flash-level WA in SSDs, yet they are also
applicable for general log-structured storage.  Take DAC \cite{chiang99} as an
example.  DAC associates each LBA with a temperature-based counter (quantified
based on the write count) and writes blocks to the segments of different
temperature levels.  Each user write promotes the LBA to a hotter segment
while each GC write demotes the LBA to a colder segment.  Other
temperature-based data placement schemes follow the similar idea of DAC.
Specifically, the above designs adopt different metrics to
measure block temperatures, such as access frequencies (in ML \cite{stoica13},
MQ \cite{yang17}, and ETI \cite{shafaei16}), recency (in FADaC
\cite{kremer19}), hotness (in SFS \cite{min12}), access counts (in DAC
\cite{chiang99}), sequentiality (in SFR \cite{yang17}), and update intervals
(in WARCIP \cite{yang19}).

\begin{figure*}[!t]
\centering
\begin{tabular}{@{\ }cc}
\includegraphics[width=3.3in]{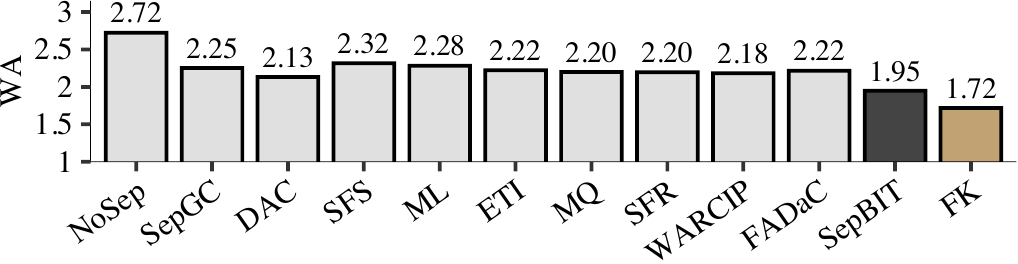} & 
\includegraphics[width=3.3in]{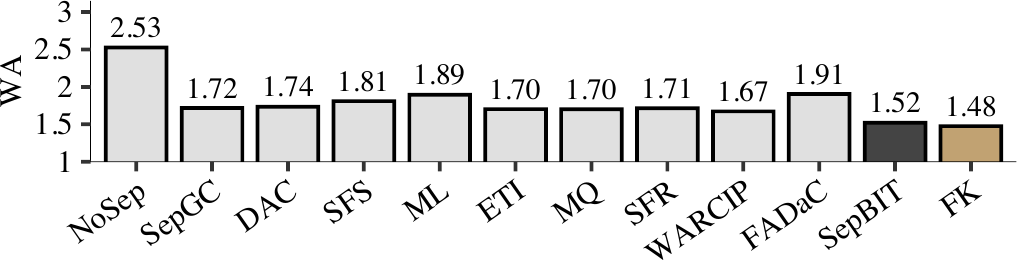}
\vspace{-3pt}\\ 
{\small (a) Overall WA of Greedy} & 
{\small (b) Overall WA of Cost-Benefit} \\
\includegraphics[width=3.3in]{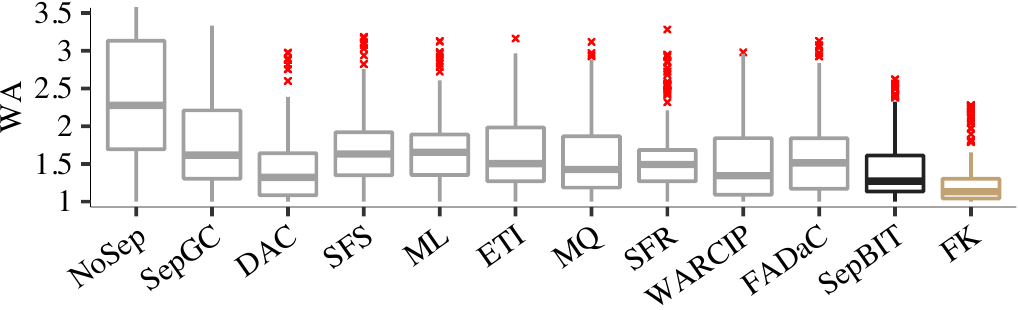} & 
\includegraphics[width=3.3in]{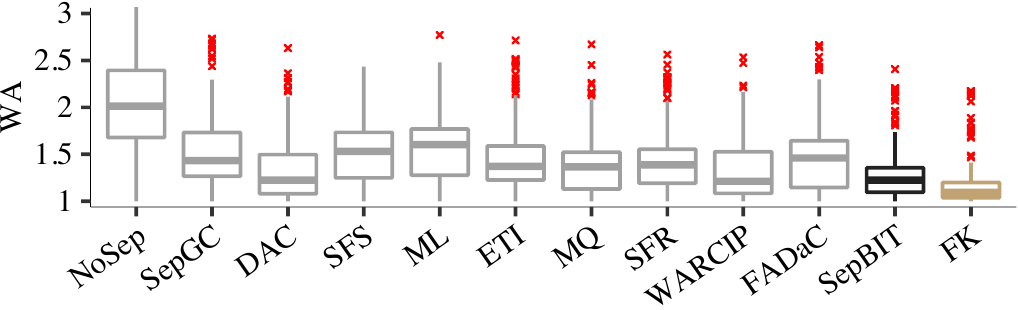} 
\vspace{-3pt}\\ 
{\small (c) Per-volume WA of Greedy} & 
{\small (d) Per-volume WA of Cost-Benefit} 
\end{tabular}
\vspace{-6pt}
\caption{Exp\#1 (Impact of segment selection). } 
\label{fig:selection}
\vspace{-6pt}
\end{figure*}

We also consider three baseline strategies.
\begin{itemize}[leftmargin=*] \itemsep=0pt \parskip=0pt
\item
{\bf NoSep} appends any written blocks (either user-written blocks or
GC-rewritten blocks) to the same
open segment. 
\item
{\bf SepGC \cite{van14}} separates written blocks by user-written blocks and
GC-rewritten blocks, and writes them into two different open segments. 
\item
{\bf Future knowledge (FK)} assumes that the BIT of each written block is
known in advance.  For a written block (either a user-written block or a
GC-rewritten block), if its invalidation will occur within $t$~bytes since the
written time, we write the block to the $\lceil \frac{t}{s} \rceil$-th open
segment, where $s$ is the segment size (in bytes).  Given the limited number
of open segments, FK uses the last open segment to store all user-written
blocks and GC-rewritten blocks if their BITs do not belong to the prior open
segments. We annotate the lifespan of each block in the traces in advance, so
that we can compute the BITs during evaluation.

\end{itemize}

Note that FK represents an {\em oracular} baseline that leverages
future knowledge for placement decisions.  It is identical to the ideal
scheme (\S\ref{subsec:ideal}) if there are unlimited memory and storage
budgets. Otherwise, with limited memory and storage budgets, it applies
future knowledge to group a subset of blocks in a limited number of
segments, and applies trivial data placement for the remaining blocks.
Thus, FK represents both the ideal data placement scheme that has no memory
and storage constraints and the trivial data placement scheme with the memory
and storage constraints; the latter serves the baseline in our experiments.

By default, we configure six classes (each containing one open segment) for
data placement for all schemes, except for NoSep, SepGC, and ETI.
For NoSep, we configure one class for all written blocks; 
for SepGC, we configure two classes, one for user-written blocks and
one for GC-rewritten blocks; for ETI, we configure two classes for
user-written blocks and one class for GC-rewritten blocks.  For MQ, SFR, and
WARCIP, as they focus on separating user-written blocks only, we configure
five classes for user-written blocks and the remaining class for GC-rewritten
blocks.  For DAC, SFS, ML, FADaC, and FK, since they do not differentiate
user-written blocks and GC-rewritten blocks, we let them use all six classes
for all written blocks. We adopt the default settings as described in the
original papers of the existing schemes.

\subsection{Results}
\label{subsec:results}

\noindent
{\bf Summary of findings.} Our major findings include:
\begin{itemize}[leftmargin=*]
\item 
\sysname achieves the lowest WA among all data placement schemes (except FK)
for different segment selection algorithms (Exp\#1), different segment sizes
(Exp\#2), and different GP thresholds (Exp\#3).  
\item
We show that \sysname provides accurate BIT inference (Exp\#4).
\item
We provide a breakdown analysis on \sysname, and show that it achieves a low WA
by separating each set of user-written blocks and GC-rewritten blocks
independently (Exp\#5).
\item
\sysname achieves the lowest WA in the Tencent Cloud traces (Exp\#6).
\item
\sysname shows high WA reduction for highly skewed workloads (Exp\#7).
\item
We provide a memory overhead analysis and show that \sysname achieves low
memory overhead for a majority of the volumes (Exp\#8).
\item
Our prototype evaluation shows that \sysname achieves the highest throughput
in a majority of the volumes (Exp\#9).
\end{itemize}

\paragraph{Default configuration.}  Our default GC policy uses Cost-Benefit
\cite{rosenblum92,rumble14} for segment selection and fixes the segment
size and the GP threshold for triggering GC as 512\,MiB and 15\%,
respectively; in Exp\#1-Exp\#3, we vary each of the configurations for
evaluation. For real-world workloads, we use the Alibaba Cloud traces except
for Exp\#5.

\paragraph{Exp\#1 (Impact of segment selection).} We compare \sysname with
existing data placement schemes using Greedy \cite{rosenblum92} and
Cost-Benefit \cite{rosenblum92,rumble14} for segment selection in GC
(\S\ref{subsec:gc}).

Figures~\ref{fig:selection}(a) and \ref{fig:selection}(b) depict the overall 
WA across all 186 volumes under Greedy and Cost-Benefit, respectively. With
separation in data placement, \sysname reduces the overall WA of NoSep by
28.5\% and 39.8\% under Greedy and Cost-Benefit, respectively.  More
importantly, \sysname achieves the lowest WA compared with all existing data
placement schemes (except FK).  It reduces the overall WA of SepGC and the
eight state-of-the-art data placement schemes (i.e., excluding NoSep and FK)
by 8.6-15.9\% and 9.1-20.2\% under Greedy and Cost-Benefit, respectively. 
Compared with FK, the overall WA of \sysname is 13.5\% and 3.1\% higher under
Greedy and Cost-Benefit, respectively.  In short, \sysname is highly efficient
in WA mitigation under real-world workloads. Note that some data placement
schemes even show a higher WA than SepGC, which performs simple separation of
user-written blocks and GC-rewritten blocks, mainly because they fail to
effectively group blocks with similar BITs 
(\S\ref{subsec:motivation}). 

Figures~\ref{fig:selection}(c) and \ref{fig:selection}(d) show the
boxplots of per-volume WAs over all 186 volumes under Greedy and Cost-Benefit,
respectively (we omit outliers of NoSep with very high WAs).  \sysname has
the lowest 75th percentiles (1.61 and 1.36) among all existing data placement
schemes (except FK) under Greedy and Cost-Benefit, while the second lowest one
is DAC (1.64 and 1.50), respectively.  This shows that \sysname effectively
reduces WAs in individual volumes with diverse workloads.  In
particular, Cost-Benefit is more effective in the WA reduction of \sysname
than Greedy, as the gap of the 75th percentiles between \sysname and the second
lowest one increases from 1.8\% in Greedy to 9.4\% in Cost-Benefit.  Compared
with FK, for 75th percentiles, \sysname has 23.6\% and 12.9\% higher WA under
Greedy and Cost-Benefit, respectively.

\paragraph{Exp\#2 (Impact of segment sizes).}  We vary the segment size from
64\,MiB to 512\,MiB.  For fair comparisons, we fix the amount of data (both
valid and invalid data) to be retrieved in each GC operation as 512\,MiB,
meaning that a GC operation collects eight, four, two, and one segment(s) for
the segment sizes of 64\,MiB, 128\,MiB, 256\,MiB, and 512\,MiB, respectively.
We focus on comparing NoSep, SepGC, WARCIP, \sysname, and FK, as they show the
lowest WAs among existing data placement for various segment sizes. 
We present the complete results in Appendix~\ref{sec:app_eval}.

\begin{figure*}[t]
\begin{tabular}{@{\ }ccc}
  \multicolumn{2}{c}{
    \includegraphics[width=3in]{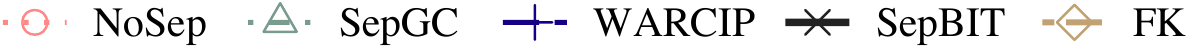}
    \vspace{-12pt}
  } \\
  \parbox[t]{2.2in}{
    \includegraphics[width=2.2in]{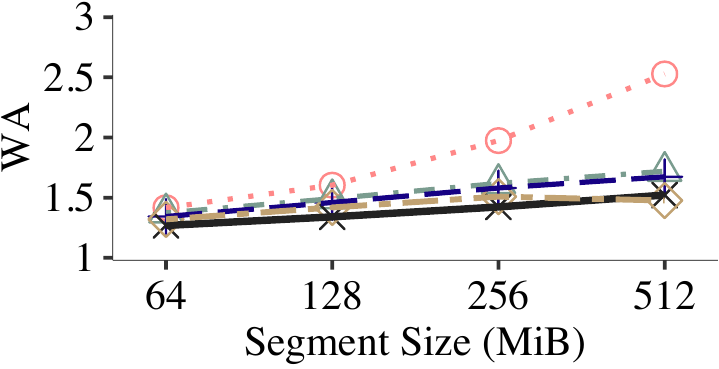}
    \caption{Exp\#2 (Impact of segment sizes). }
    \label{fig:overall_segsize}
  } &
  \parbox[t]{2.2in}{
    \centering
    \includegraphics[width=2.2in]{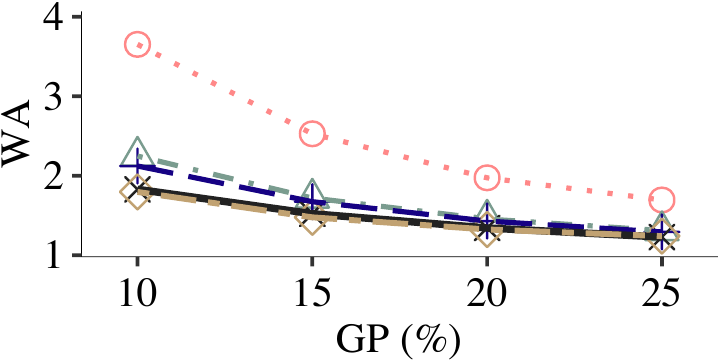}
    \caption{Exp\#3 (Impact of GP thresholds). }
    \label{fig:overall_gp}
  } &
  \parbox[t]{2.2in}{
    \centering
    \hspace{-3pt}
    \includegraphics[width=2.2in]{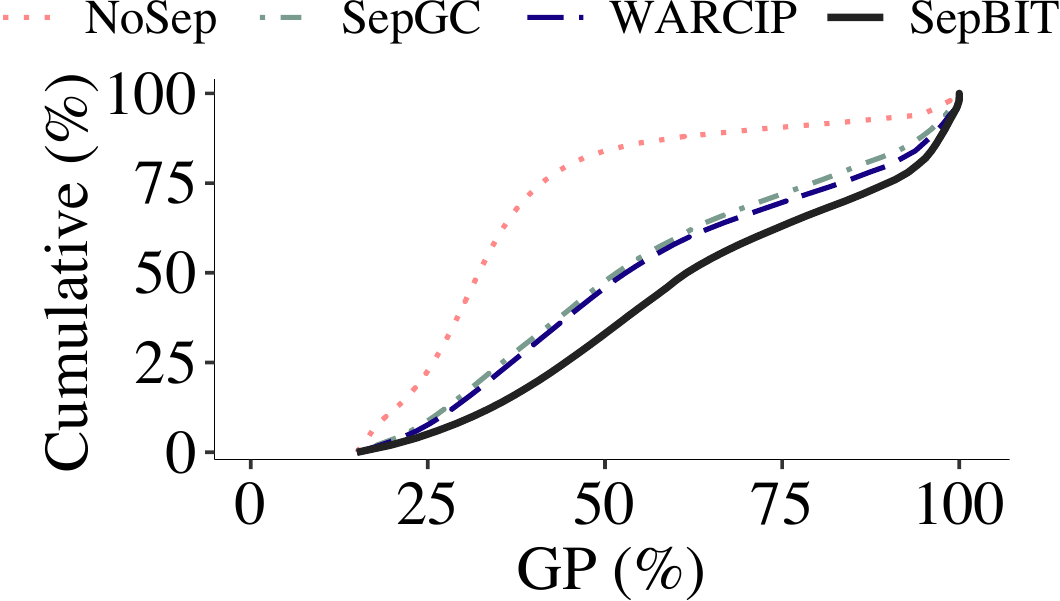}
    \caption{Exp\#4 (BIT inference analysis). }
    \label{fig:predict}
  } 
\end{tabular}
\vspace{-3pt}
\end{figure*}

Figures~\ref{fig:overall_segsize} depicts the overall WA versus the segment
size. Overall, using a smaller segment size yields a lower WA, as a GC
operation can perform more fine-grained selection of segments for more
efficient space reclamation.
Again, \sysname achieves the lowest WA compared with all existing data placement
schemes; for example, its WAs are 5.5\%, 8.2\%, and 10.0\% lower than WARCIP
for the segment sizes of 64\,MiB, 128\,MiB, and 256\,MiB, respectively.
Interestingly, \sysname even has a lower WA (by 3.9-5.7\%) than FK when the
segment size is in the range of 64\,MiB to 256\,MiB. The reason is that FK
currently groups blocks of close BITs in five open segments, while the last
open segment stores all blocks (we now configure six classes in total)
(\S\ref{subsec:schemes}).  If the segment size is smaller, FK can only
group fewer blocks in the limited number of open segments, so it becomes less
effective of grouping blocks of close BITs.

\paragraph{Exp\#3 (Impact of GP thresholds).} We vary the GP thresholds from
10\% to 25\%.  We again focus on comparing the overall WAs of NoSep, SepGC,
WARCIP, \sysname, and FK as in Exp\#2. Figure~\ref{fig:overall_gp} shows the
overall WA versus the GP threshold.  A larger GP threshold has a lower WA in
general, as it is easier for a GC operation to select segments with high GPs.
\sysname still shows the lowest WA. It has 5.0-13.8\% lower WAs than WARCIP for
different GP thresholds.  Compared with FK, \sysname has comparable WAs with
differences smaller than 1.8\%, for different GP thresholds.

\paragraph{Exp\#4 (BIT inference analysis).}
We study the effectiveness of the BIT inference in \sysname.  Note that
\sysname does not explicitly compute the estimated BIT of a block, but instead
assigns blocks into classes corresponding to different ranges of estimated
BITs (\S\ref{subsec:impl}).  To examine the effectiveness of BIT inference,
our intuition is that each valid block that is rewritten during GC indicates
that we incorrectly infer its BIT and places it into an incorrect segment.
Thus, we can examine the GP of each collected segment to estimate the
inference accuracy, such that a higher GP implies more accurate inference.
We use the Cost-Benefit selection algorithm and fix the segment size and GP for
triggering GC as 512\,MiB and 15\%, respectively. We study NoSep, SepGC,
WARCIP, and \sysname (WARCIP has the second lowest WA). We aggregate the
collected segments during GC for all 186 volumes. 

Figure~\ref{fig:predict} depicts the cumulative distributions of collected
segments across different GPs for different schemes. The median GPs of the
collected segments for NoSep, SepGC, WARCIP, and \sysname are 32.3\%, 51.6\%,
52.9\%, and 61.5\%, respectively.  \sysname has the highest GPs, implying that
it also has the highest accuracy in inferring BITs.  WARCIP only shows a
slightly higher GP of the collected segments than SepGC, so its WA reduction
over SepGC is marginal.

\paragraph{Exp\#5 (Breakdown analysis).} We analyze how different components
of \sysname contribute to WA reduction.  Recall that \sysname separates
written blocks into the user-written blocks and GC-rewritten blocks, and further
separates each set of user-written blocks and GC-rewritten blocks
independently.  In our analysis, we consider NoSep (i.e., without
separation), SepGC (i.e., separating written blocks into the user-written
blocks and GC-rewritten blocks), and two variants:
\begin{itemize}[leftmargin=*]  
\item 
{\em UW}: It further separates user-written blocks based on SepGC, but without 
separating GC-rewritten blocks.  It maintains three classes: Classes~1 and 2
store short-lived blocks and long-lived blocks as in \sysname, respectively,
while Class~3 stores all GC-rewritten blocks.
\item 
{\em GW}: It further separates GC-rewritten blocks based on SepGC, but without
separating user-written blocks.  It maintains four classes: Class~1 stores all
user-written blocks, and Classes~2-4 store GC-rewritten blocks as in
Classes~4-6 of \sysname. 
\end{itemize}

\begin{figure}[t]
\centering
\begin{tabular}{@{\ }c@{\ }c}
\includegraphics[width=1.65in]{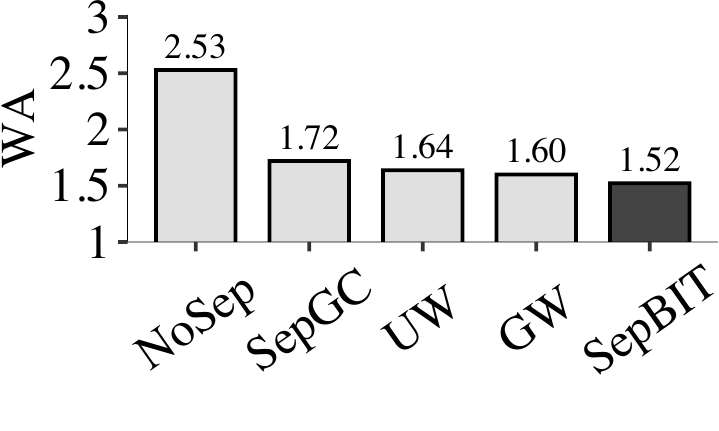} & 
\includegraphics[width=1.65in]{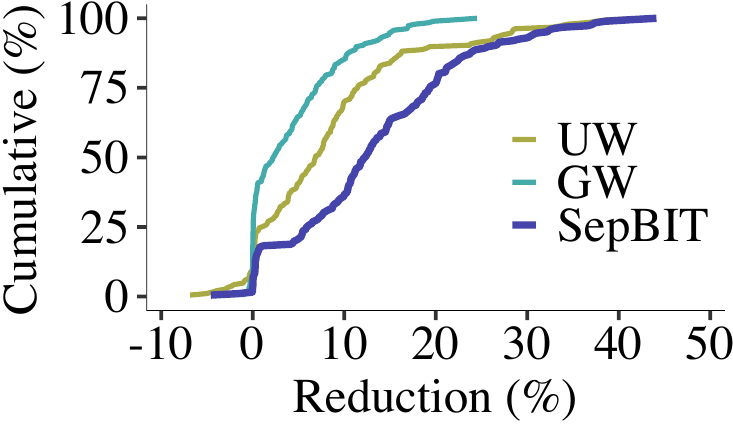} 
\vspace{-3pt}\\ 
{\small (a) Overall WA} & 
{\small (b) Per-volume WA reduction}
\end{tabular}
\vspace{-6pt}
\caption{Exp\#5 (Breakdown analysis).} 
\label{fig:breakdown}
\vspace{-6pt}
\end{figure}

\begin{figure*}[t]
\centering
\begin{tabular}{@{\ }cc}
\includegraphics[width=3.3in]{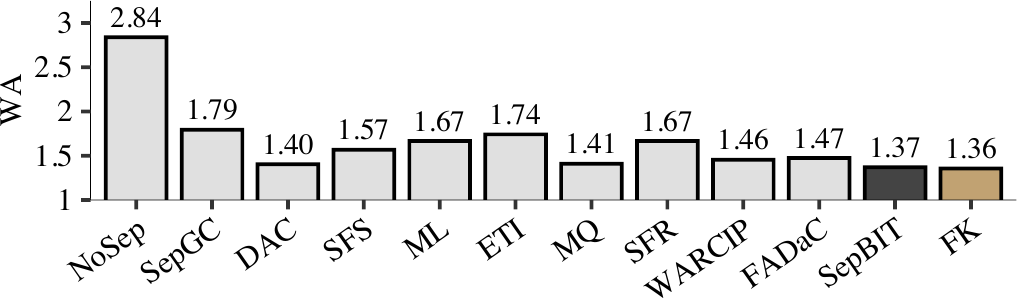} & 
\includegraphics[width=3.3in]{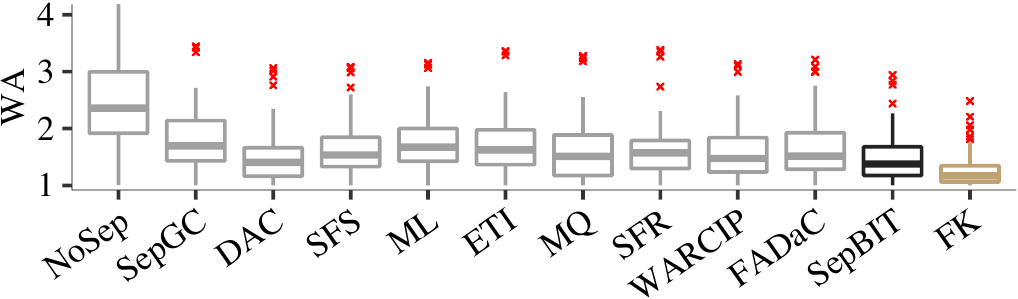} 
\vspace{-3pt}\\ 
{\small (a) Overall WA} & 
{\small (b) Per-volume WA} 
\end{tabular}
\vspace{-6pt}
\caption{Exp\#6 (Results on the Tencent Cloud traces). } 
\label{fig:tencent}
\vspace{-6pt}
\end{figure*}

Figure~\ref{fig:breakdown}(a) shows the overall WAs of different data
placement schemes.  UW and GW reduce WA by 35.2\% and 36.7\% compared with
NoSep, respectively; they also reduce WA by 4.8\% and 7.0\% compared with
SepGC, respectively.  The findings show that more fine-grained separation of
each set of user-written blocks and GC-rewritten blocks brings further WA
reduction.  Also, \sysname reduces WA by 7.0\% and 4.9\% compared with UW and
GW, respectively, meaning that \sysname can combine the benefits of UW and GW.

Figure~\ref{fig:breakdown}(b) further shows the cumulative distributions of the
WA reductions of UW, GW, and \sysname compared with SepGC across all volumes.
UW, GW, and \sysname can reduce the WA of most of the volumes. The 75th
percentiles of reductions of UW and GW are 11.4\% and 6.9\%, respectively, and
their highest WA reductions are 43.3\% and 24.5\%, respectively. By combining
UW and GW, the 75th percentile of the WA reductions of \sysname
compared with SepGC improves to 19.3\% with the highest WA reduction as 44.1\%.

\paragraph{Exp\#6 (Results on the Tencent Cloud traces).}
We validate the effectiveness of \sysname on the Tencent Cloud traces
\cite{zhang20osca}. We pre-process the traces the same as for the Alibaba Cloud
traces (\S\ref{subsec:traces}) and select 271 out of 4,995 volumes. We run all
the schemes as in Exp\#1, using Cost-Benefit for segment selection and fixing
the segment size and the GP threshold as 512\,MiB and 15\%, respectively.

Figure~\ref{fig:tencent} depicts the overall WA
and the per-volume WA across all 271 volumes. Among all existing
data placement schemes, \sysname achieves the lowest overall WA.  Its overall
WA is 2.5-21.3\% lower than those of the eight existing schemes and
1.1\% higher than that of FK. Compared with the second lowest scheme DAC,
\sysname has similar 50th and 75th percentiles of per-volume WA, and reduces
the 90th percentile of per-volume WA from 2.09 to 1.97.

\paragraph{Exp\#7 (Impact of workload skewness).} We study how \sysname works
in workloads of different skewness. We set the selection algorithm as Greedy
instead of Cost-Benefit, since Cost-Benefit also leverages the workload skewness
to reduce WA and we want to exclude its impact from our analysis.

We inspect the skewness of each volume in the Alibaba Cloud traces,
and analyze the correlation between the per-volume skewness and the WA
reduction percentage of \sysname over NoSep. We also present the results for
synthetic workloads in Appendix~\ref{sec:app_eval}.
Since not all real-world workloads have good fitness to a Zipf distribution
\cite{yang16}, we
describe the per-volume skewness according to how write traffic aggregates in
the most frequently updated blocks. Specifically, we compute the percentage of
aggregated write traffic over the top 20\% frequently written blocks. To
show the relationship between the percentage of aggregated writes and the
skewness factor of the Zipf distribution, Table~\ref{tab:zipf20} shows the
percentage of write traffic over the top 20\% frequently written blocks and
the corresponding skewness factor $\alpha$; note that the numbers are
generated using 10\,GiB of write WSS.

\begin{table}[t]
\small
\centering
\begin{tabular}{c|c|c|c|c|c|c}
  \hline
  {\bf Skewness $\alpha$} & 0 & 0.2 & 0.4 & 0.6 & 0.8 & 1             \\ 
  \hline
  {\bf Pct. (\%)} & 20 & 27.6 & 38.1 & 52.4 & 71.1 & 89.5           \\ 
  \hline
\end{tabular}
\vspace{-3pt}
\caption{The percentage of write traffic over top-20\% blocks in Zipf workloads of different skewness.}
\label{tab:zipf20}
\vspace{-6pt}
\end{table}
  
\begin{figure}[t]
\begin{tabular}{@{\ }cc}
    \parbox[t]{1.6in}{
      \includegraphics[width=1.6in]{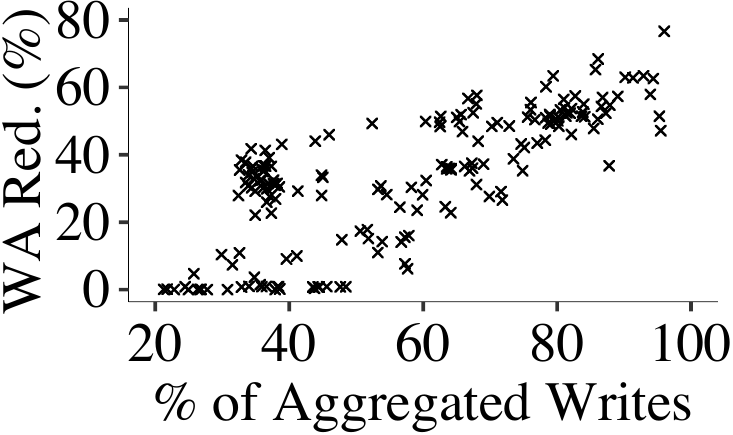} 
      \vspace{-18pt}
      \caption{Exp\#7 (Impact of workload skewness).}
      \label{fig:skewness}
    }
    &
    \parbox[t]{1.6in}{
      \includegraphics[width=1.6in]{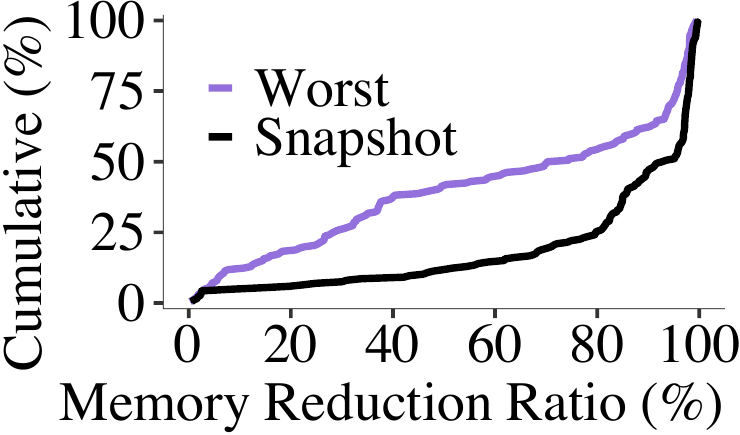} 
      \vspace{-18pt}
      \caption{Exp\#8 (Memory overhead). }
      \label{fig:memory}
    }
\end{tabular}
\vspace{-6pt}
\end{figure}

Figure~\ref{fig:skewness} shows the results. Each point represents one volume.
The x-axis is the percentage of aggregated write traffic over top 20\%
frequently written blocks and the y-axis is the WA reduction of \sysname over
NoSep. We see a positive correlation between the percentage of aggregated
write traffic and the WA reduction (we also find that the p-value is smaller
than 0.01 for the Pearson correlation coefficient 0.75, meaning that the
positive correlation is statistically significant).
For the volumes with percentages of aggregated write traffic larger than 80\%,
\sysname reduces the WA by at least 38.0\% with a maximum reduction of 76.7\%. 

\paragraph{Exp\#8 (Memory overhead analysis).}
We analyze the memory overhead of \sysname using the Alibaba Cloud traces.
Recall that \sysname tracks only the unique LBAs inside the FIFO queue
(\S\ref{subsec:impl}), instead of maintaining the mappings for all LBAs in the
write working set.  We report the memory overhead reduction of \sysname as one
minus the ratio of the number of unique LBAs in the FIFO queue to the number of
unique LBAs in the write working set.  To quantify the reduction, for each
volume, we collect all values of the number of unique LBAs in the FIFO queue at
runtime when $\ell$ (\S\ref{subsec:impl}) is updated.  To avoid bias due to the
cold start of trace replay, for each volume, we exclude the beginning 10\% of
the values. We also collect the number of unique LBAs at the end of the traces.
We consider two cases, namely (i) the worst case and (ii) the snapshot case. In
the worst case, we use the maximum number of unique LBAs in the FIFO queue for
all volumes; it assumes that each volume has its peak number of unique LBAs in
the FIFO queue and incurs the most memory. In the snapshot case, we use the
number of unique LBAs at the end of the traces, representing a snapshot of the
system status.

From our analysis, we find that in the worst case, \sysname reduces the overall
memory overhead by 44.8\%, while in the snapshot case, \sysname reduces the
overall memory overhead by 71.8\%. To calculate the actual memory overhead,
suppose that the mapping for each LBA has 8~bytes, in which both
the LBA and the FIFO position are of size 4~bytes each (a 4-byte LBA can
represent an address space of $2^{32} \times 2^{12} = 16$\,TiB for 4-KiB
blocks). Since the
aggregated write WSS of the 186 volumes is 20.3\,TiB (\S\ref{subsec:traces}),
\sysname reduces the overall memory overhead from $20.3 \cdot
\frac{2^{40}}{2^{12}} \cdot 8 = 41.6$\,GiB to $41.6 \cdot (1-71.8\%) =
11.7$\,GiB.

Figure~\ref{fig:memory} further depicts the cumulative distributions of the
memory overhead reductions across volumes under both the worst case and the
snapshot case. In the worst case, \sysname reduces the memory overhead by more
than 72.3\% in half of the volumes and the highest memory overhead reduction is
99.5\%; in the snapshot case, the median reduction is 93.1\% with the highest
reduction as 99.7\%. In the snapshot case, the 25th, 50th, and
75th percentiles of the number of unique LBAs across volumes are 99\,K,
 1,063\,K, and 6,190\,K, respectively, 
while the 25th, 50th, 75th percentiles of the number of total LBAs in the FIFO
queue across volumes are 398\,K, 2,242\,K, and 8,857\,K, respectively.
The reason of the differences among volumes is their different degrees of
skewness. The volumes with higher skewness see more aggregated traffic
patterns, and hence the number of recently updated LBAs is much smaller
compared with the write WSS. 

\paragraph{Exp\#9 (Prototype evaluation).} We deploy our log-structured block
storage system prototype (\S\ref{subsec:impl}) on a machine equipped with an
Intel Xeon Silver 4215 CPU, 96\,GiB DDR4 RAM, and 4$\times$128\,GiB Intel
Optane Persistent Memory modules.  The machine runs Ubuntu 20.04.2 LTS with
kernel 5.4.0.  

Due to the limited storage capacity in our testbed machine, we focus on 20
volumes whose write traffic ranks the top 31-50 among the 186 volumes in the
Alibaba Cloud traces.  Their write traffic ranges from 0.82\,TiB to 2.82\,TiB,
and their WAs under NoSep range from 1.00 to 4.96.  Specifically, 9 volumes
have their WAs less than 1.1, while 7 volumes have their WAs greater than 3.0.

Also, our evaluation rate-limits user writes while GC is running due to the
capacity constraint.  The reason is that a GC operation removes the invalid
blocks only after rewriting all valid blocks.  If we issue user writes at full
speed while GC is running, the storage space may run out.  Thus, we limit the
rate of user writes as 40\,MiB/s while GC is running; otherwise, we issue user
writes at full speed.  We measure the {\em write throughput} (i.e., the number
of user-written bytes divided by the total time for replaying each volume).

We compare \sysname with NoSep, DAC, and WARCIP, based on our previous
experiments that DAC and WARCIP perform the best among existing schemes and
NoSep serves as the baseline.  We configure the segment selection algorithm,
the segment size, and the GP threshold as Cost-Benefit, 512\,MiB, and 15\%,
respectively.

\begin{figure}[t]
\centering
\begin{tabular}{@{\ }c@{\ }c}
\includegraphics[width=1.65in]{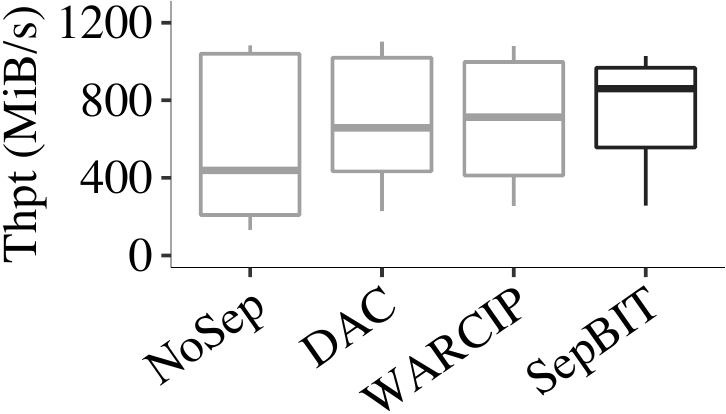} & 
\includegraphics[width=1.65in]{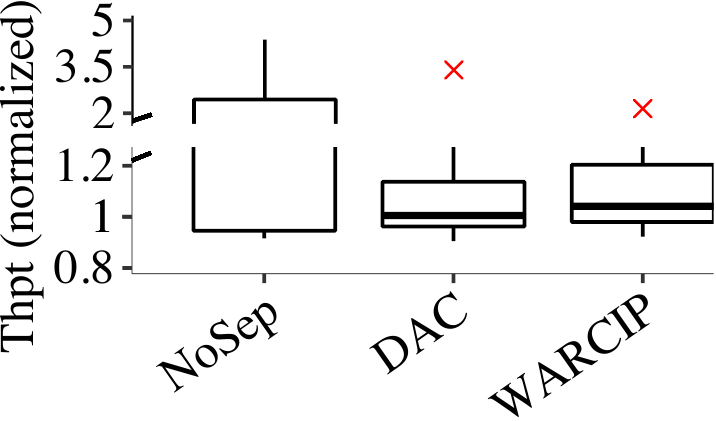} \\
{\small (a) Absolute throughput} & 
{\small (b) Normalized throughput} 
\end{tabular}
\vspace{-6pt}
\caption{Exp\#9 (Prototype evaluation).} 
\label{fig:prototype}
\vspace{-6pt}
\end{figure}

Figures~\ref{fig:prototype}(a) and \ref{fig:prototype}(b) show the boxplots of
the absolute write throughput and the normalized write throughput of \sysname
(w.r.t.  NoSep, DAC, and WARCIP) in individual volumes for different schemes,
respectively.  \sysname achieves the highest throughput for the 25th and 50th
percentiles, at 556.1\,MiB/s and 859.4\,MiB/s, which are 28.3\% and 20.4\%
higher than the second best, respectively. 

For the 75th percentile, the absolute throughput of \sysname is 6.9\%, 5.2\%,
and 3.0\% lower than those of NoSep, DAC, and WARCIP, respectively
(Figure~\ref{fig:prototype}(a)).  The reason is that such volumes (with
top-25\% throughput) have low WAs (less than 1.1) and hence are less affected
by GC. Compared with other schemes, \sysname spends extra time to access the
FIFO queue (\S\ref{subsec:impl}) and has slightly degraded throughput. 

\section{Related Work}
\label{sec:related}

\noindent
{\bf GC in SSDs.}  We evaluated several existing data placement schemes
(\S\ref{subsec:schemes}) for mitigating the WA of flash-level GC in SSDs.
Other data placement schemes build on the use of program contexts
\cite{kim19} or the prediction of block temperature based on neural networks
\cite{yang19ml}.  Some empirical studies evaluate the data placement
algorithms on an SSD platform \cite{lee13}, or characterize how real-world I/O
workloads affect GC performance \cite{yadgar21}.  In particular, Yadgar {\em et al.}
\cite{yadgar21} also investigate the impact of the number of separated classes
in data placement based on the temperature-based data scheme MultiLog
\cite{stoica13}. In contrast, \sysname builds on the BIT for data placement,
backed by the empirical studies from real-world I/O traces.  ML-DT
\cite{chakraborttii21} uses neural networks to predict the block death time.
Compared with ML-DT, \sysname infers BITs only with the last user write time in
a simpler manner.

Besides data placement, existing studies propose segment selection algorithms
to reduce the WA of flash-level GC.  In addition to Greedy and Cost-Benefit
(\S\ref{subsec:gc}), Cost-Age-Times \cite{chiang99cat} considers the
cleaning cost, data age, and flash erasure counts in segment selection.
Windowed Greedy \cite{hu09}, Random-Greedy \cite{li13}, and d-choices
\cite{van13} are variants of Greedy in segment selection. Desnoyers
\cite{desnoyers14} models the WA of different segment selection algorithms and
hot-cold data separation. \sysname can work in conjunction with those
algorithms.

\paragraph{GC in file systems.}  Several studies examine the GC performance
for log-structured file systems.  Matthew {\em et al.} \cite{matthews97} improve
the GC performance by adapting GC to the system and workload behaviors.  SFS
\cite{min12} separates blocks by hotness (i.e., write frequency divided by
age). Some studies reduce WA using file system semantics in data placement;
for example, WOLF \cite{wang02} groups blocks by files or directories, while
hFS \cite{zhang07} and F2FS \cite{lee15} separate data and metadata.
Extending \sysname with file system awareness is a future work. 

\paragraph{GC for RAID and distributed storage.} Some studies address the GC
performance issues in RAID and distributed storage, such as reducing the WA of
Log-RAID systems \cite{chiueh14} and mitigating the interference between GC
and user writes via GC scheduling in RAID arrays \cite{shin13,kim19}.
RAMCloud \cite{rumble14} targets persistent distributed in-memory storage. It
proposes two-level cleaning to maximize memory utilization by coordinating 
GC operations in memory and disk backends.  It also corrects the original
Cost-Benefit algorithm \cite{rosenblum92} for accurate segment selection. 
Our work focuses on data placement for WA mitigation and is orthogonal to
those studies.

\section{Conclusion}

We propose \sysname, a novel data placement scheme that mitigates WA caused by
GC in log-structured storage by grouping blocks with similar estimated BITs.
Inspired from the ideal data placement that minimizes WA (i.e., WA$=$1) using
future knowledge of BITs, \sysname leverages the skewed write patterns of
real-world workloads to infer BITs. It separates written blocks into
user-written blocks and GC-rewritten blocks and performs fine-grained
separation in each set of user-written blocks and GC-rewritten blocks. To
group blocks with similar BITs, it infers the BITs of user-written blocks and
GC-rewritten blocks by estimating their lifespans and residual lifespans,
respectively.  Evaluation on production traces shows that \sysname achieves
the lowest WA compared with eight state-of-the-art data placement schemes.

\paragraph{Acknowledgements.}
We thank our shepherd, Keith Smith, and the anonymous reviewers for their
comments. This work was supported in part by Alibaba Group via the Alibaba
Innovation Research (AIR) program. The corresponding author is Patrick P. C.
Lee.

\bibliographystyle{abbrv}
\bibliography{reference}

\clearpage
\newpage

\appendix

\section*{Appendix}

In Appendix~\ref{sec:app_tencent}, we provide all the analysis results for the
Tencent Cloud traces.  In Appendix~\ref{sec:app_derivation}, we provide the
derivation of the conditional probability for inferring BITs
(\S\ref{subsec:userwrite} and \S\ref{subsec:gcwrite}).  In
Appendix~\ref{sec:app_eval}, we provide additional evaluation results,
including the complete trace analysis results for the Alibaba Cloud traces,
sensitivity analysis of \sysname with regard to the varying number of classes
and varying separation thresholds, and the impact of workload skewness on
\sysname using synthetic workloads.

\begin{figure*}[!t]
\begin{tabular}{@{\ }ccc}
\parbox[t]{2.2in}{
\centering
\includegraphics[width=2.1in]{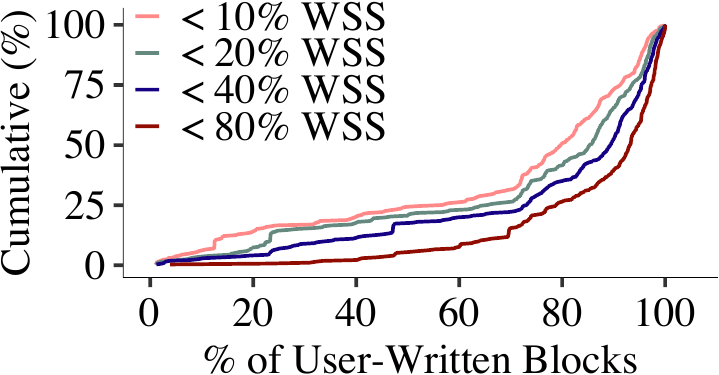}
\vspace{-6pt}
\caption{Percentages of user-written blocks with different short lifespans.}
\label{fig:o1_tencent}
} 
& 
\parbox[t]{2.2in}{
\centering
\includegraphics[width=2.1in]{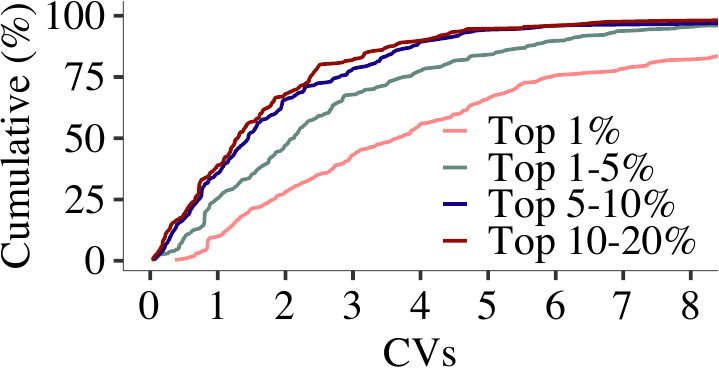}
\vspace{-6pt}
\caption{CVs of the lifespans of frequently updated blocks with different
similar update frequencies.}
\label{fig:o2_tencent}
} 
&
\parbox[t]{2.2in}{
\centering
\includegraphics[width=2.1in]{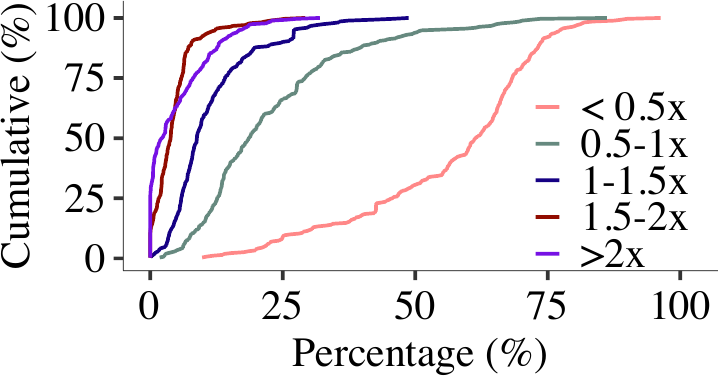} 
\vspace{-6pt}
\caption{Percentages of rarely updated blocks with different lifespans.}
\label{fig:o3_tencent}
} 
\end{tabular}
\end{figure*}

\begin{figure*}[!t]
\centering
\begin{tabular}{@{\ }cc}
\parbox[t]{3.32in}{
\centering
\includegraphics[width=3.3in]{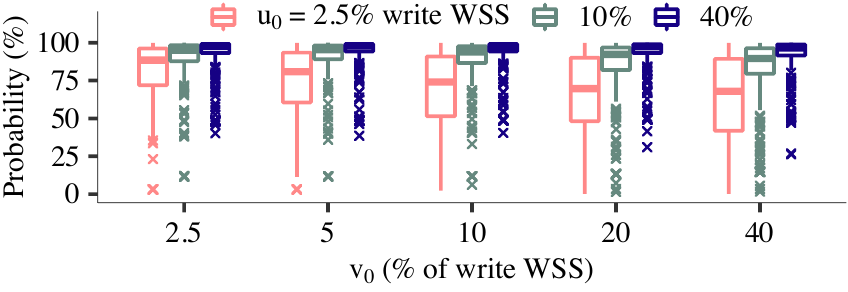}
\vspace{-12pt}
\caption{Inferring BITs of user-written blocks: Boxplots of 
$\Pr(u \le u_0 \mid v \le v_0)$ for different $u_0$ and $v_0$ in Tencent Cloud 
workloads.}
\label{fig:tencent_userwrite}
} 
& 
\parbox[t]{3.32in}{
\centering
\includegraphics[width=3.3in]{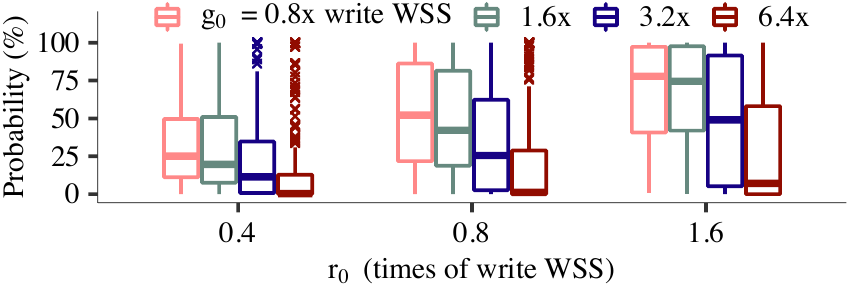}
\vspace{-12pt}
\caption{Inferring BITs of GC-rewritten blocks: Boxplots of 
$\Pr(u \le g_0 + r_0 \mid u \ge g_0)$
for different $r_0$ and $g_0$ in Tencent Cloud workloads.}
\label{fig:tencent_gcwrite}
} 
\end{tabular}
\end{figure*}

\section{Analysis for Tencent Cloud Traces}
\label{sec:app_tencent}

We analyze the Tencent Cloud traces \cite{zhang20osca} to validate our
observations with the Alibaba Cloud traces.  Our analysis includes
Observations~1-3 (\S\ref{subsec:motivation}) as well as the inferring of BITs
for user-written blocks (\S\ref{subsec:userwrite}) and GC-rewritten blocks
(\S\ref{subsec:gcwrite}).  We focus on in total selected 271 volumes that have
their write WSS larger than 10\,GiB and write traffic exceeding 2$\times$
their write WSS.

In Observation 1, we examine the percentages of user-written blocks that fall
into different lifespan range groups with short lifespans that are smaller than
write WSS for each volume. Figure~\ref{fig:o1_tencent} shows the cumulative
distribution of the percentage of user-written blocks across all volumes in
different lifespan groups. We find that half of the volumes have more than
93.2\% of their user-written blocks with lifespans shorter than 80\% of their
WSSes, and have more than 79.8\% user-written blocks with lifespans shorter
than only 10\% WSSes.

In Observation 2, we examine the coefficient of variance (CV) of the lifespans
of the blocks in different groups of frequently updated blocks, i.e., top 1\%, top
1-5\%, top 5-10\%, top 10-20\%. Figure~\ref{fig:o2_tencent} shows the
cumulative distribution of CVs across all volumes (49, 13, 9, 5 volumes in the
four groups have CVs exceeding 8, respectively). Similar to what we see in 
the Alibaba Cloud traces, 25\% of the volumes have their CVs exceeding 5.93, 3.78,
2.81, 2.37 in the four groups, respectively.

In Observation 3, we examine lifespan distributions of rarely updated blocks
that are updated no more than four times. In half of the volumes, more than
75.8\% of their write working sets contain rarely updated blocks. We again
divide the rarely updated blocks into five groups that are partitioned by the
lifespans of 0.5$\times$, 1$\times$, 1.5$\times$, and 2$\times$ of their write
WSSes. Figure~\ref{fig:o3_tencent} shows the cumulative distributions
of the percentages of rarely updated blocks in different groups of lifespans
across all volumes. In 25\% of the volumes, more than 76.2\% of the rarely
updated blocks have their lifespans shorter than 0.5$\times$ and the medians 
for the remaining four groups are 16.7\%, 6.8\%, 3.1\%, 1.0\%, presenting
similar results as the Alibaba Cloud traces.

To infer the lifespans of user-written blocks, we define a conditional
probability $\Pr(u \le u_0 \mid v \le v_0)$ (\S\ref{subsec:userwrite}).  Same
as in \S\ref{subsec:userwrite}, we set the values of both $v_0$ and $u_0$ as
different percentages of write WSS to examine the probability.
Figure~\ref{fig:tencent_userwrite} depicts the boxplots of the conditional
probabilities over all volumes for different $u_0$ and $v_0$. The conditional
probability remains high in most of the volumes. We see similar results as in
the Alibaba Cloud traces. For $v_0$ that equals to 40\% of write WSS, the medians
of the conditional probabilities are in the range of 67.9-96.5\%, and the 75th
percentiles are in the range of 89.3-98.9\%.

To infer the lifespans of GC-rewritten blocks, we define a conditional
probability $\Pr(u \le g_0 + r_0 \mid u \ge g_0)$ (\S\ref{subsec:gcwrite}).
Same as in \S\ref{subsec:gcwrite}, we set both $r_0$ and $g_0$ as different
percentages of the write WSS.  Figure~\ref{fig:tencent_gcwrite} depicts the
boxplots of the conditional probabilities over all volumes for different $g_0$
and $r_0$. Again, we see similar results as in the Alibaba Cloud traces.  For a
fixed $r_0$, the conditional probabilities have significant differences for
varying $g_0$. For example, if we fix $r_0$ as $1.6\times$ of write WSS, and
$g_0$ increases from $0.8\times$ to $6.4\times$ of WSS, the median of
probabilities drop from 77.9\% to 7.1\%.

\section{Derivation of the Conditional Probability for Inferring BITs}
\label{sec:app_derivation}

We present the detailed derivation of the conditional probabilities for
inferring BITs of user-written blocks (\S\ref{subsec:userwrite}) and
GC-rewritten blocks (\S\ref{subsec:gcwrite}).

According to the definitions in \S\ref{subsec:userwrite}, for user-written
blocks, we investigate
\begin{align*}
  \Pr(u \le u_0 \mid v \le v_0) 
  = \frac{\Pr(u \le u_0 \text{ and } v \le v_0)} {\Pr(v \le v_0)}, 
\end{align*}
where the denominator is expressed as:
\begin{align*} 
  \Pr(v \le v_0) 
  & = \sum\nolimits_{i=1}^{n} \Pr(v \le v_0 \mid A_{b'} = i) \cdot
  \Pr(A_{b'} = i) \\
  & = \sum\nolimits_{i=1}^{n} (1 - (1 - p_i)^{v_0}) \cdot p_i,
\end{align*}
and the numerator is expressed as: 
\begin{align*} 
  & \Pr(u \le u_0 \text{ and } v \le v_0) \\
  = & \sum\nolimits_{i=1}^{n} \Pr(u \le u_0 \text{ and } v \le v_0 \mid 
  A_{b} = i) \cdot \Pr(A_{b} = i) \\  
  = & \sum\nolimits_{i=1}^{n} (1 - (1 - p_i)^{u_0}) \cdot (1 - (1 - p_i)^{v_0}) \cdot p_i.
\end{align*}

According to the definitions in \S\ref{subsec:gcwrite}, for GC-wewritten
blocks, we investigate
\begin{align*}
  \Pr(u \le g_0 + r_0 \mid u \ge g_0) 
  & = \frac{\Pr(g_0 \le u \le g_0 + r_0)}{\Pr(u \ge g_0)},
\end{align*}
where the numerator and denominator correspond to all user-written blocks whose
lifespans range from $g_0 \le u \le g_0 + r_0$ and $u \ge g_0$, respectively.
The
numerator and denominator are respectively:
\begin{align*}
  \Pr(g_0 \le u \le g_0 + r_0) & = \sum\nolimits_{i=1}^{n} p_i \cdot ((1-p_i)^{g_0} - (1-p_i)^{g_0 + r_0}) \\
  \text{ and }
  \Pr(u \ge g_0) & = \sum\nolimits_{i=1}^{n} p_i \cdot (1-p_i)^{g_0}.
\end{align*}

\section{Additional Evaluation Results}
\label{sec:app_eval}

Recall that in \S\ref{sec:eval}, we investigate the impact of segment sizes
(Exp\#2) and the GP thresholds (Exp\#3), but only show some representative
results. We now present the complete results in Exp\#A1 and Exp\#A2. We also
conduct sensitivity analysis for \sysname with regard to the varying number of
classes and the separation thresholds in Exp\#A3 and Exp\#A4, respectively.
Finally, we study the impact of workload skewness to \sysname using synthetic
workloads in Exp\#A5.

\begin{figure}[!t]
\centering
\begin{tabular}{@{\ }c}
\includegraphics[width=3in]{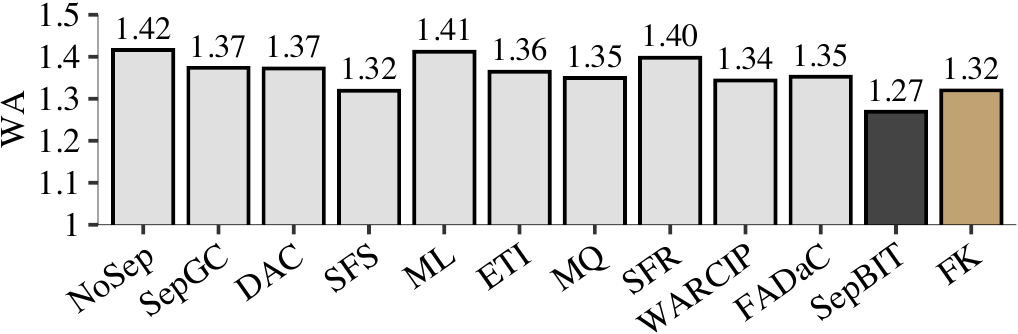} 
\vspace{-3pt}\\ 
{\small (a) Segment size of 64\,MiB} \\ 
\includegraphics[width=3in]{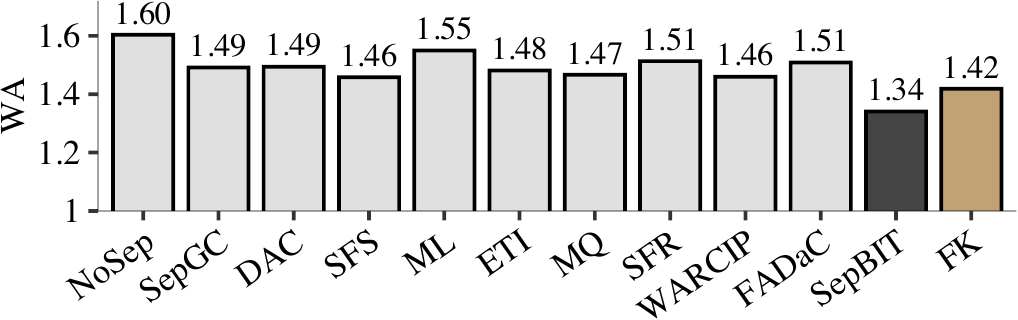} 
\vspace{-3pt}\\ 
{\small (b) Segment size of 128\,MiB} \\ 
\includegraphics[width=3in]{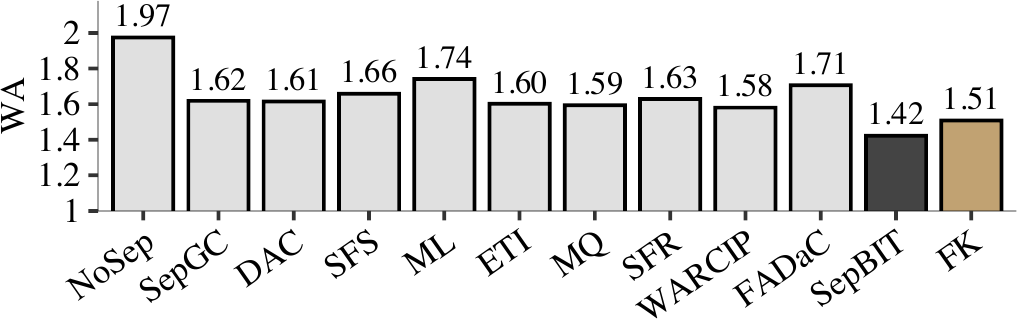} 
\vspace{-3pt}\\ 
{\small (c) Segment size of 256\,MiB} 
\end{tabular}
\vspace{-6pt}
\caption{Exp\#A1 (Impact of segment sizes). Overall WA.} 
\label{fig:appendix_overall_segsize}
\vspace{-6pt}
\end{figure}

\begin{figure}[!t]
\centering
\begin{tabular}{@{\ }c}
\includegraphics[width=3in]{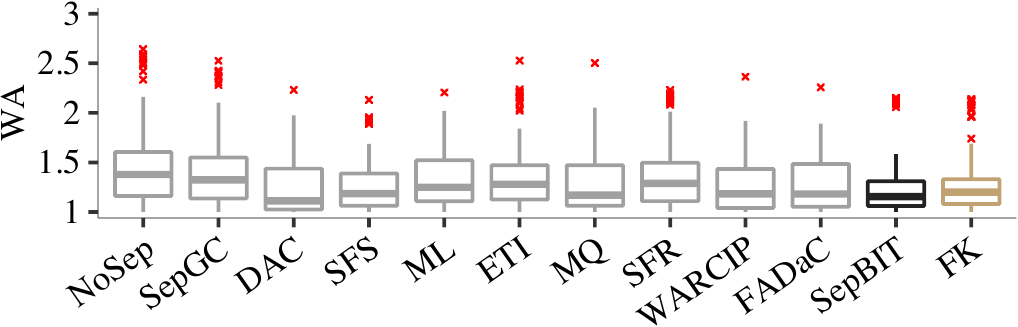} 
\vspace{-3pt}\\ 
{\small (a) Segment size of 64\,MiB} \\ 
\includegraphics[width=3in]{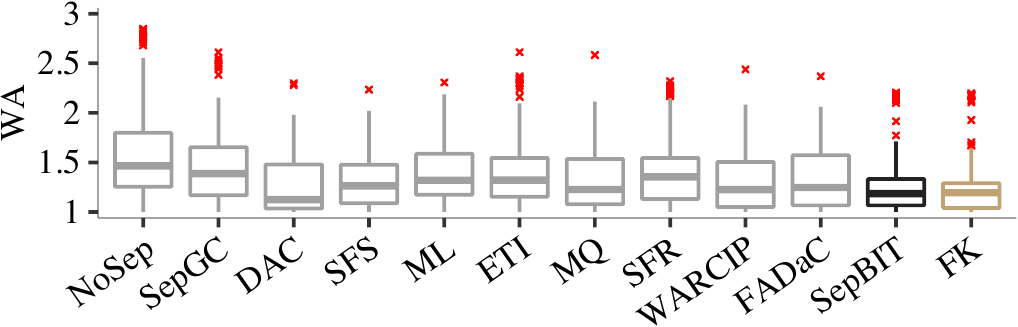} 
\vspace{-3pt}\\ 
{\small (b) Segment size of 128\,MiB} \\ 
\includegraphics[width=3in]{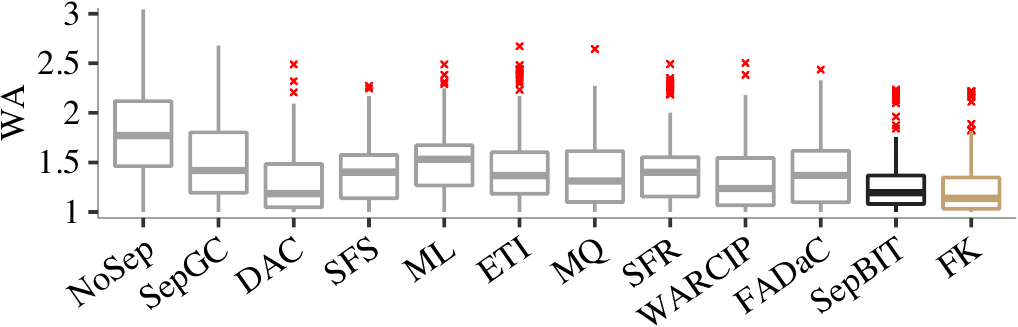} 
\vspace{-3pt}\\ 
{\small (c) Segment size of 256\,MiB} 
\end{tabular}
\vspace{-6pt}
\caption{Exp\#A1 (Impact of segment sizes). Per-volume WA.} 
\label{fig:appendix_pervolume_segsize}
\vspace{-6pt}
\end{figure}

\paragraph{Exp\#A1 (Impact of segment sizes). }
We first show the complete results of the overall WA versus different segment
sizes in Figure~\ref{fig:appendix_overall_segsize}, in addition to the five
algorithms shown in Exp\#2 (\S\ref{sec:eval}).
Figures~\ref{fig:appendix_pervolume_segsize}(a),
\ref{fig:appendix_pervolume_segsize}(b), and
\ref{fig:appendix_pervolume_segsize}(c) further depict the boxplots of the
single volume WA over all volumes under segment sizes of 64\,MiB, 128\,MiB, and
256\,MiB, respectively (note that we have already shown the case of
512\,MiB segment sizes in Exp\#1). We omit 9 outliers in NoSep whose WA is
above 3 in Figure~\ref{fig:appendix_pervolume_segsize}(c). For three segment
sizes, \sysname still achieves the lowest 75th percentiles (1.31, 1.35, and
1.37, respectively), with 5.8\%, 8.8\% and 7.4\% lower than the second lowest
ones (1.39 in SFS, 1.48 in SFS, and 1.48 in DAC), respectively. Compared with
FK, for the 75th percentiles, \sysname has comparable WA, with 1.5\% lower,
1.5\% higher and 4.6\% higher compared with FK, under segment sizes of
64\,MiB, 128\,MiB and 256\,MiB, respectively.

\begin{figure}[!t]
\centering
\begin{tabular}{@{\ }c}
\includegraphics[width=3in]{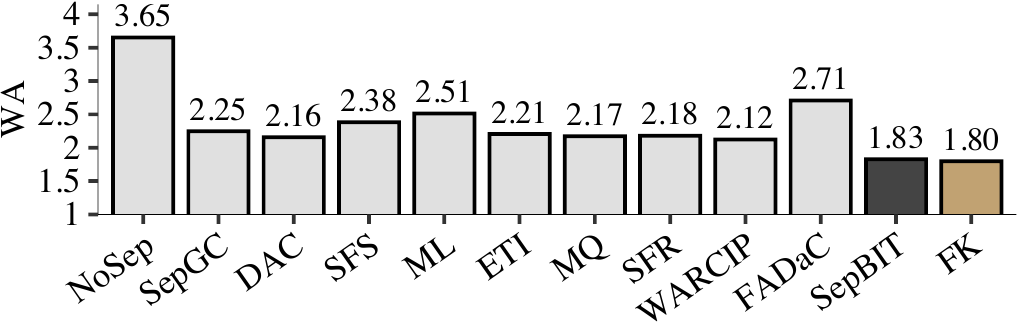} 
\vspace{-3pt}\\ 
{\small (a) GP threshold of 10\%} \\ 
\includegraphics[width=3in]{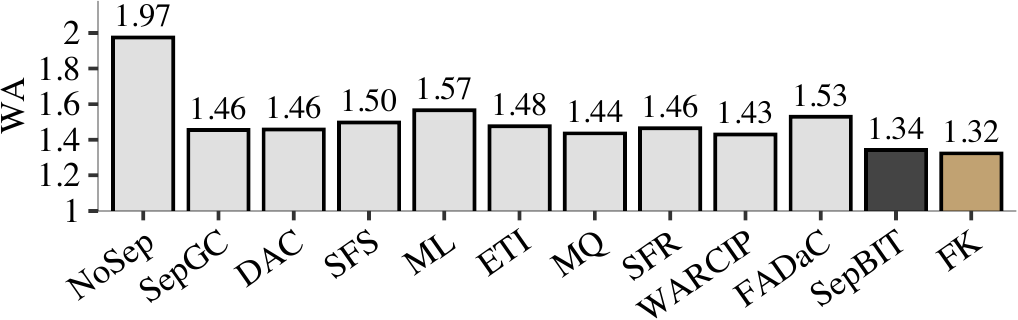} 
\vspace{-3pt}\\ 
{\small (b) GP threshold of 20\%} \\ 
\includegraphics[width=3in]{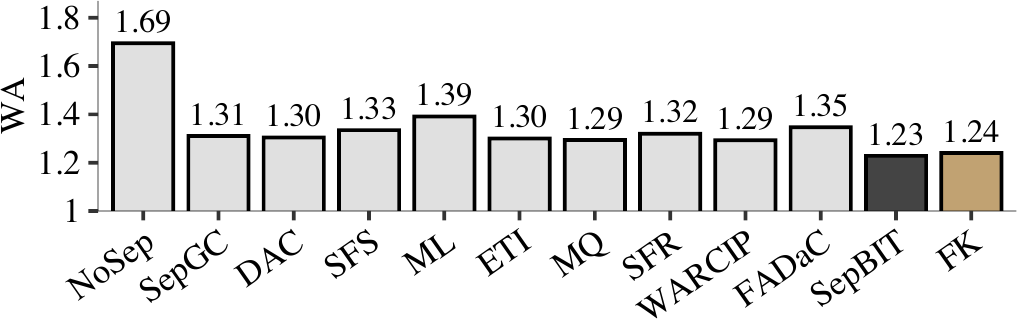} 
\vspace{-3pt}\\ 
{\small (c) GP threshold of 25\%} 
\end{tabular}
\vspace{-6pt}
\caption{Exp\#A2 (Impact of GP thresholds). Overall WA.} 
\label{fig:appendix_overall_gp}
\vspace{-6pt}
\end{figure}

\begin{figure}[!t]
\centering
\begin{tabular}{@{\ }c}
\includegraphics[width=3in]{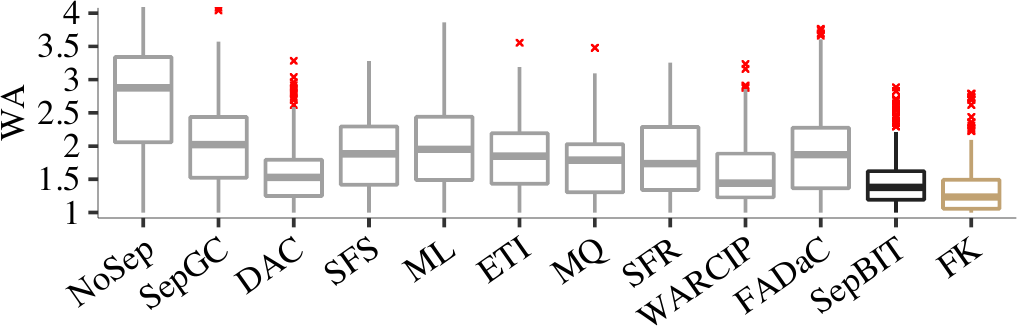} 
\vspace{-3pt}\\ 
{\small (a) GP threshold of 10\%} \\ 
\includegraphics[width=3in]{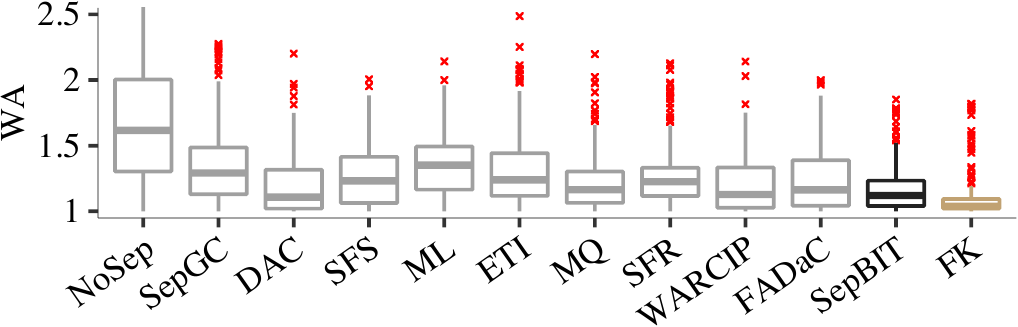} 
\vspace{-3pt}\\ 
{\small (b) GP threshold of 20\%} \\ 
\includegraphics[width=3in]{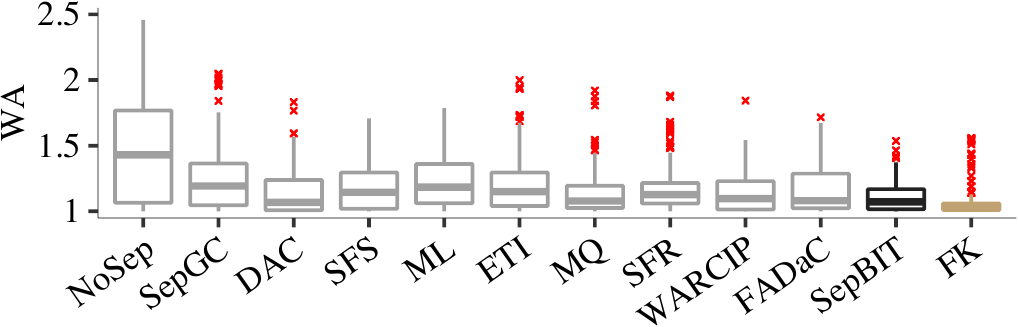} 
\vspace{-3pt}\\ 
{\small (c) GP threshold of 25\%} 
\end{tabular}
\vspace{-6pt}
\caption{Exp\#A2 (Impact of GP thresholds). Per-volume WA.} 
\label{fig:appendix_pervolume_gp}
\vspace{-6pt}
\end{figure}

\paragraph{Exp\#A2 (Impact of GP thresholds). }
We first show the complete results of the overall WA versus different GP
thresholds in Figure~\ref{fig:appendix_overall_gp}, in addition to the five
algorithms shown in Exp\#3 (\S\ref{sec:eval}).
Figures~\ref{fig:appendix_pervolume_gp}(a), \ref{fig:appendix_pervolume_gp}(b)
and \ref{fig:appendix_pervolume_gp}(c) further depict the boxplots of single
volume WA over all volumes under GP thresholds of 10\%, 20\% and 25\% GP,
respectively. We omit 31 outliers and 8 outliers whose WA is above 4, in NoSep
and SepGC, respectively in Figure~\ref{fig:appendix_pervolume_gp}(a). Also, we
omit 16 outliers in NoSep whose WA is above 2.5 in
Figure~\ref{fig:appendix_pervolume_gp}(b) (note that
we already show the case of 15\% in Exp\#1 (\S\ref{sec:eval})). Configuring a
higher percentage of the GP threshold generally achieves lower WA. Compared
with existing data placement schemes, \sysname reduces the 75th percentiles by
9.2\%, 4.2\%, and 1.7\% upon the second smallest schemes (1.80 in DAC, 1.31 in
MQ, and 1.20 in MQ) under the GP thresholds of 10\%, 20\% and 25\%,
respectively.  Compared with FK, \sysname still has higher WA at 75th
percentiles, by 9.4\%, 14.7\% and 11.3\% higher under the GP thresholds of
10\%, 20\% and 25\%, respectively.



\begin{figure}[t]
\centering
\begin{tabular}{@{\ }c@{\ }c}
\includegraphics[width=1.65in]{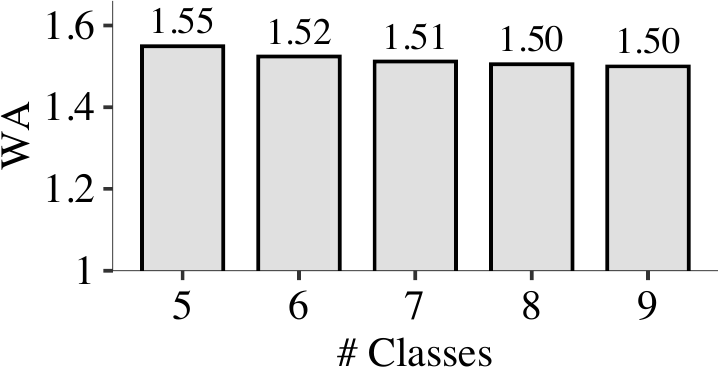} & 
\includegraphics[width=1.65in]{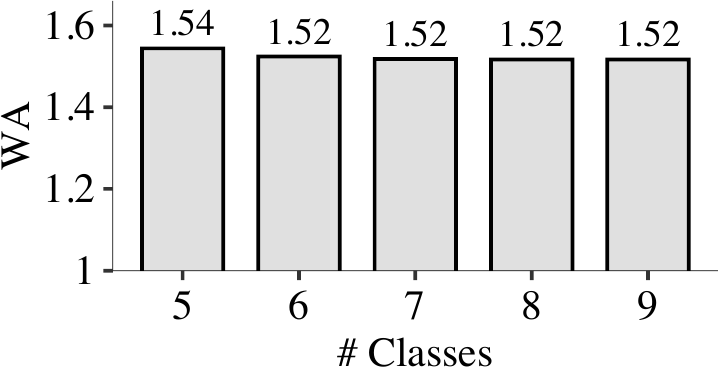} 
\vspace{-3pt}\\
{\small (a) Method 1} & 
{\small (b) Method 2}  
\end{tabular}
\vspace{-6pt}
\caption{Exp\#A3 (Sensitivity analysis of varying number of classes). } 
\vspace{-10pt}
\label{fig:openseg}
\end{figure}

\begin{figure}[t]
\centering
\begin{tabular}{@{\ }c@{\ }c}
\includegraphics[width=1.65in]{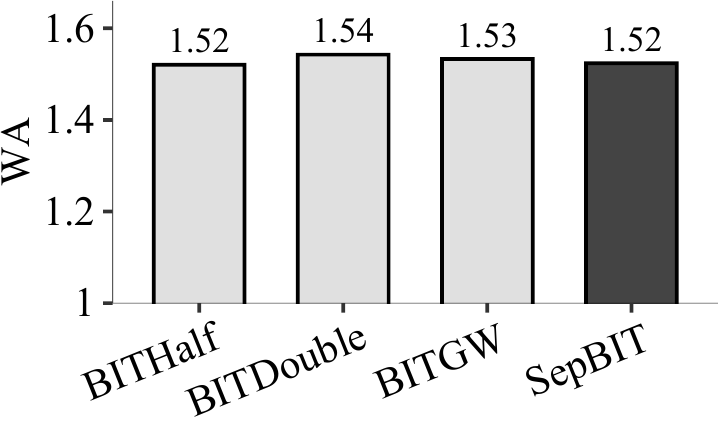} & 
\includegraphics[width=1.65in]{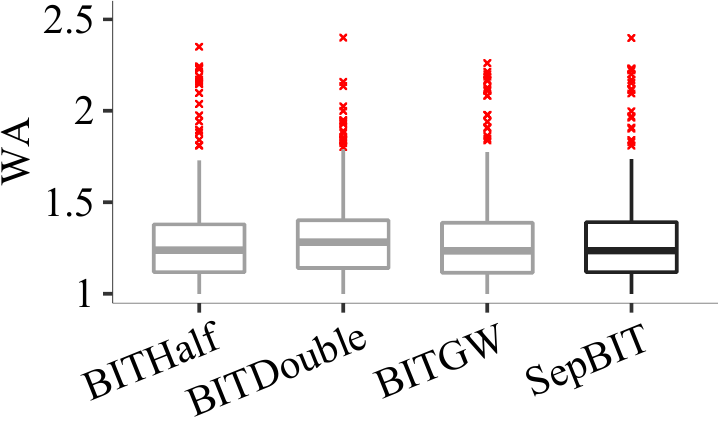} \\ 
{\small (a) Overall WA} & 
{\small (b) Per-volume WA}  
\end{tabular}
\vspace{-6pt}
\caption{Exp\#A4 (Sensitivity analysis of varying separation thresholds).} 
\label{fig:thresholds}
\vspace{-6pt}
\end{figure}

\begin{figure}[t]
\centering
\begin{tabular}{@{\ }c}
\includegraphics[width=2.2in]{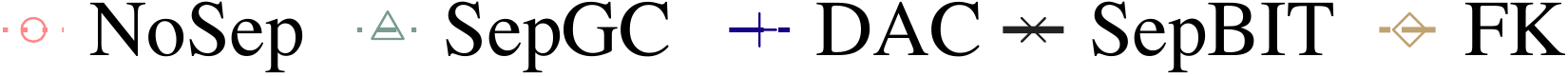} \\
\includegraphics[width=2.2in]{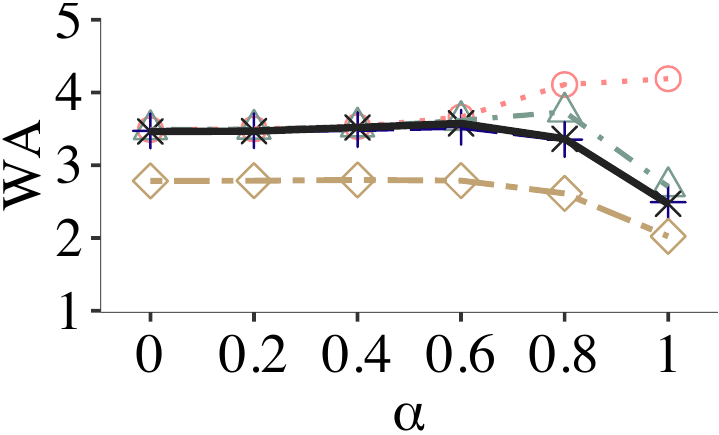} 
\end{tabular}
\vspace{-6pt}
\caption{Exp\#A5 (Impact of workload skewness).}
\label{fig:app_skewness}
\vspace{-9pt}
\end{figure}

\paragraph{Exp\#A3 (Sensitivity analysis of varying number of classes).} We
study how the WA of \sysname varies across the number of classes.  We start
with the default configuration with the following settings: (i) We separate
user-written blocks to Classes~1-2 and GC-rewritten blocks to the remaining
classes; (ii) For GC-rewritten blocks from Class~1, we append them to Class~3;
(iii) The thresholds are adaptive according to $\ell$ (\S\ref{subsec:impl}).
Then we use two methods to change the age thresholds for the remaining classes
of the GC-rewritten blocks.  For Method~1, we configure the age thresholds as
$16^{i/(c-1)} \times \ell$ for Class~$(3+i)$, where $c$ is the number of the
remaining classes, and $i=1, 2,\cdots, c-1$.  For example, for $c=3$ remaining
classes (our default), we set the age thresholds as $4\ell$ and $16\ell$; for
$c=5$ remaining classes, we set the age thresholds as $2\ell$, $4\ell$,
$8\ell$, and $16\ell$, which separate the remaining GC-rewritten blocks into
Classes~4-8.  For Method~2, we configure the age threshold as
$4^{i} \times \ell$ for Class~$(3+i)$. As a result, for $c=2$ remaining
classes, there is only one threshold $4\ell$, and for $c=5$ remaining
classes, the age thresholds are $4\ell$, $16\ell$, $64\ell$, and $256\ell$,
which separates the remaining GC-rewritten blocks into Classes~4-8.

Figure~\ref{fig:openseg} shows the overall WA of both methods, where the total
number of classes varies from 5 to 9. The overall WA slightly decreases when
the number of classes increases from 5 to 9 for both methods, from 1.55 to
1.50 and from 1.54 to 1.52, respectively. The results of Method~1 indicate
that \sysname already achieves high effectiveness; more fine-grained
separations only increase resource usage of open segments with negligible WA
reductions. For Method~2, the reason of minor reductions is
that the blocks whose ages exceed $16\ell$ are unlikely to be updated in the
whole trace period and thus do not need further separations.

\paragraph{Exp\#A4 (Sensitivity analysis of varying separation thresholds).} We
study how the WA of \sysname varies on different choices of separation
thresholds for separating user-written blocks and GC-rewritten blocks. We
consider three variants based on \sysname: 
\begin{itemize}[leftmargin=*]
\item 
{\em BITHalf}: It halves all thresholds for separation.
\item 
{\em BITDouble}: It doubles all thresholds for separation.
\item 
{\em BITGW}: It sets the age thresholds for separating GC-rewritten blocks to
Classes~4-6 as $\ell_{3}$ and $\ell_{3} + \ell_{4}$, where $\ell_{3}$ and
$\ell_{4}$ refer to the average lifespans of collected segments in Classes~3
and 4, respectively. The rationale is to take into account the residual
lifespan patterns of GC-rewritten blocks.
\end{itemize}

Figure~\ref{fig:thresholds} depicts the overall
WA and boxplots of per-volume WAs. The overall WA has no
significant differences. All schemes, including \sysname, achieve a similar
overall WA at around 1.52.  They also have similar percentiles among all
volumes, in which the differences of each percentile are less than 0.03 across
all schemes.  Thus, \sysname is insensitive to the choices of separation
thresholds. 

\paragraph{Exp\#A5 (Impact of workload skewness).} We study how \sysname works in workloads of different skewness.  We fix the selection algorithm as Greedy
instead of Cost-Benefit, since Cost-Benefit also leverages the skewness in
segment selection to reduce WA and we want to exclude its impact from our
analysis.  We fix the segment size as 512\,MiB and the GP threshold as 15\%.

We first synthesize workloads such that the probabilities of being written in
each request for the LBAs in the working sets follow a Zipf distribution
(\S\ref{subsec:userwrite}). Specifically, we set the write WSS as 50\,GiB and
vary the skewness factor $\alpha$ from 0 to 1. We generate 3\,TiB of write
traffic for each skewness factor. To reproduce the high variance in frequently
updated blocks as in Observation~2 (\S\ref{subsec:motivation}), we divide the
50\,GiB working set into two regions: a frequently updated region (10\,GiB)
and a rarely updated region (40\,GiB). We change the write probabilities of
LBAs in the frequently updated region for every 512\,MiB written blocks, while
keeping the overall Zipf distribution unchanged. We show the results of NoSep,
SepGC, \sysname, FK, and DAC since DAC has the lowest WA among all eight
existing schemes for different skewness.

Figure~\ref{fig:app_skewness} depicts the WA versus different skewness factors
for different schemes. For NoSep, the WA grows with the workload skewness, due
to the mixing of frequently and rarely updated blocks.  With ideal future
knowledge, FK achieves WA reduction for workloads of all different skewness.
The effectiveness of all existing data placement schemes (including other
schemes not shown) and \sysname increases with the workload skewness (i.e., a
larger $\alpha$), and both DAC and \sysname achieve the lowest WAs (with
smaller than 2\% differences).  In particular, when $\alpha=0.8$ and $1$,
\sysname reduces the WA by 18.1\% and 41.2\% compared to NoSep, respectively.

\end{document}